\documentclass[10pt,twocolumn,journal]{IEEEtran}
\usepackage{latexsym}
\usepackage{amsfonts}
\usepackage{amsbsy}
\usepackage{amsmath,amssymb}
\usepackage{times}
\usepackage{graphicx}
\usepackage{enumerate}
\usepackage[usenames]{color}
\usepackage[dvips]{pstcol}
\usepackage{epstopdf}

\usepackage{subeqnarray}
\usepackage{cases}
\usepackage{stfloats}
\usepackage{cases}
\usepackage{subeqnarray}
\usepackage{amsmath}
\usepackage{multicol}
\usepackage{multirow}
\usepackage{bigstrut}
\usepackage{amsthm}
\usepackage{booktabs}
\usepackage{algorithm}
\usepackage{algorithmic}
\usepackage{bm}
\usepackage{amsthm}
\usepackage{cite}
\input epsf
\title{Intelligent Reflecting Surface Enhanced Wireless Network: Two-Timescale Beamforming Optimization}
\author{Ming-Min Zhao, \IEEEmembership{Member,~IEEE,} Qingqing Wu, \IEEEmembership{Member,~IEEE,} Min-Jian Zhao, \IEEEmembership{Member,~IEEE,} and Rui Zhang, \IEEEmembership{Fellow,~IEEE}
	\thanks{
		
		This article was presented in part at the IEEE 11th Sensor Array and Multichannel Signal Processing Workshop (SAM) 2020 \cite{Zhao2020SAM}.
		
		M. M. Zhao and M. J. Zhao are with the College of Information Science and Electronic Engineering, Zhejiang University, Hangzhou, China 310027 (email: \{zmmblack, mjzhao\}@zju.edu.cn). Q. Wu is with the State Key Laboratory of Internet of Things for Smart City and Department of Electrical and Computer Engineering, University of Macau, Macau, China 999078 (email: qingqingwu@um.edu.mo). R. Zhang is with the Department of Electrical and Computer Engineering, National University of Singapore, Singapore 117583 (email: elezhang@nus.edu.sg).
	}
}

\begin{document}
		\maketitle
	\begin{abstract} 
Intelligent reflecting surface (IRS) has drawn a lot of attention recently as a promising new solution to achieve high spectral and energy efficiency for future wireless networks. By utilizing massive low-cost passive reflecting elements, the wireless propagation environment becomes controllable and thus can be made favorable for improving the communication performance. Prior works on IRS mainly rely on the instantaneous channel state information (I-CSI), which, however, is practically difficult to obtain for IRS-associated links due to its passive operation and large number of reflecting elements. To overcome this difficulty, we propose in this paper a new two-timescale (TTS) transmission protocol to maximize the achievable average sum-rate for an IRS-aided multiuser system under the general correlated Rician channel model. Specifically, the passive IRS phase shifts are first optimized based on the statistical CSI (S-CSI) of all links, which varies much slowly as compared to their I-CSI; while the transmit beamforming/precoding vectors at the access point (AP) are then designed to cater to the I-CSI of the users' effective fading channels with the optimized IRS phase shifts, thus significantly reducing the channel training overhead and passive beamforming design complexity over the existing schemes based on the I-CSI of all channels. Besides, for ease of practical implementation, we consider discrete phase shifts at each reflecting element of the IRS. For the single-user case, an efficient penalty dual decomposition (PDD)-based algorithm is proposed, where the IRS phase shifts are updated in parallel to reduce the computational time. For the  multiuser case, we propose a general TTS stochastic successive convex approximation (SSCA) algorithm by constructing a quadratic surrogate of the objective function, which cannot be explicitly expressed in closed-form. Simulation results are presented to validate the effectiveness of our proposed algorithms and evaluate the impact of S-CSI and channel correlation on the system performance.
	\end{abstract}

\section{Introduction}
Massive multiple-input multiple-output (MIMO) technology can achieve high spectral efficiency for wireless communication by exploiting highly directional beamforming and spatial multiplexing gains \cite{Marzetta2010}. However, equipping a large number of antennas may lead to more circuit energy consumption and higher hardware cost, especially as the wireless system evolves into the new era of  millimeter-wave (mmWave) communications \cite{Zhang2017survey}. Recently, intelligent reflecting surface (IRS) (also known as reconfigurable intelligent surface (RIS) and so on) has been proposed as a new solution to achieve high spectral efficiency with low energy and hardware cost \cite{Wu2019Magazine, Wu2018_journal, wu2019beamforming, Huang2019,Renzo2019, Basar2019}. Specifically, IRS is a passive array composed of a large number of passive reflecting elements, which can induce phase shift and/or amplitude change of the incident signal independently, thus collaboratively creating a favorable wireless signal propagation environment to enhance the communication performance. In addition, since such passive elements do not require any transmit radio frequency (RF) chains, their energy and hardware cost is much lower as compared to that of the traditional active antennas at the base stations (BSs), access points (APs), and relays. As a result, they can be densely deployed in wireless networks with a scalable cost, and yet without causing any interference to each other provided that they are deployed sufficiently far apart. Moreover, it is practically easy to integrate IRSs into the existing cellular or WiFi systems as there is no need to modify their existing infrastructure and operating standards \cite{Wu2019Magazine}. All the above advantages make IRS a promising technology for future wireless systems, particularly for indoor/hot-spot coverage and cell-edge performance enhancement.
 
IRS has been investigated recently in various aspects and under different setups, such as passive beamforming designs \cite{Yu2019, guo2019weighted, Han2019, zhang2019capacity}, IRS-aided orthogonal frequency division multiplexing (OFDM) system \cite{Yang2019, zheng2019intelligent}, IRS-aided mmWave communications \cite{cao2019intelligent}, physical layer security \cite{Cui2019, Chen2019access, xu2019resource, guan2019intelligent}, wireless power transfer \cite{Mishra2019ICASSP, Wu2019SWIPT, pan2019intelligent, WuJSAC2019}, and so on. Particularly, \cite{Wu2018_journal} showed that IRS is able to create a ``signal hot spot'' in its vicinity with an asymptotic power gain in the order of $N^2$, where $N$ denotes the number of IRS reflecting elements. Moreover, \cite{wu2019beamforming} further showed that even with practical discrete phase shifters at the IRS, the same squared power gain of $N^2$ is achievable with only a constant power loss in dB depending on the number of phase-shift levels at each reflecting element, which becomes negligible as $N$ becomes very large. Therefore, significant performance gains can be achieved with IRS as compared to conventional wireless systems without using IRS.

To fully realize the potentials of IRS-aided wireless systems, accurate channel state information (CSI) of the AP-IRS and IRS-user links are essential for optimizing the reflection coefficients. However, as the number of reflecting elements is usually very large, obtaining the accurate CSI of the AP-IRS and IRS-user links is practically difficult. As a result, how to effectively estimate the IRS-associated channels with low training/signaling overhead while still reaping most of the performance gain offered by IRS becomes a crucial issue. In the literature, there are some recent works that studied the channel estimation problem for IRS-aided wireless systems \cite{Yang2019,Mishra2019ICASSP,  zheng2019intelligent, You2019CE, He2019_CE}. Specifically, in \cite{Yang2019} and \cite{Mishra2019ICASSP}, a binary reflection controlled least-square (LS) channel estimation method was proposed, where $N$ training symbols are needed to estimate the channel coefficients associated with the IRS. In \cite{Yang2019}, an IRS element-grouping method was also proposed to reduce the training overhead, at the cost of degraded passive beamforming performance of the IRS. The authors in \cite{zheng2019intelligent} proposed a reflection pattern based channel estimation method for an IRS-enhanced OFDM system, which was shown to have superior mean squared error (MSE) channel estimation performance than that in \cite{Yang2019}, with the same amount of training symbols. This work was further extended to the discrete phase-shift case in \cite{You2019CE}. In \cite{He2019_CE}, the low-rank structure of the massive MIMO channel was exploited and the cascaded channel estimation problem for IRS was addressed by leveraging the combined bilinear factorization and matrix completion.

It is worth pointing out that in the aforementioned studies, the beamforming vectors are mainly designed based on the instantaneous CSI (I-CSI). In practice, this approach will incur high signal processing complexity and large training/signaling overhead. Moreover, most of the existing works on IRS-aided wireless systems assume that the phase shifts of the reflecting elements can be continuously adjusted. However, discrete phase-shift controls are usually desired in practice in order to lower the implementation cost of IRS \cite{wu2019beamforming}.

To tackle the above challenges, we propose in this paper a two-timescale (TTS) joint active and passive beamforming scheme for an IRS-aided multiuser multiple-input single-output (MISO) system with practical discrete phase shifts at the IRS. In the considered system, we adopt the general correlated Rician fading channel to model the various links between the AP, IRS and users. The active precoding vectors at the AP and passive phase shifts at the IRS are jointly optimized to maximize the long-term average weighted sum-rate of the users. Moreover, in order to alleviate the high signal processing complexity and training overhead for acquiring the I-CSI, we propose a practical transmission protocol based on the measured channel statistics\footnote{For the considered correlated Rician fading channel in this paper, channel statistics refer to the deterministic components as well as the fading channel correlation matrices.} and TTS beamforming optimization. Specifically, we assume that the IRS is equipped with $N$ dedicated sensors/receiving circuits for statistical CSI (S-CSI) estimation, which is easier to implement as compared to accurately tracking the I-CSI at the IRS that varies much faster than its S-CSI. Once the S-CSI is estimated and fed back to the AP, the AP performs the optimization of the IRS {\it long-term} phase shifts based on it and sends their values to the IRS, which sets the phase shifts accordingly for the subsequent time slots regardless of the instantaneous channel variations, as long as the S-CSI remains unchanged (e.g., in the case of a quasi-static user in the vicinity of the IRS). In the meanwhile, at each time slot, the {\it short-term} transmit precoding vectors at the AP are dynamically designed to cater to the effective I-CSI with fixed IRS phase shifts.

In particular, we first consider the single-user case for the purpose of exposition and drawing useful insights. By deriving an upper bound of the achievable average rate, we show that the original stochastic optimization problem can be transformed into a deterministic non-convex optimization problem. To tackle this new problem, instead of resorting to the commonly used semidefinite relaxation (SDR) method \cite{Wu2018_journal} or the successive refinement algorithm based on the block coordinate descent (BCD) method \cite{wu2019beamforming}, we propose a new algorithm by leveraging the penalty dual decomposition (PDD) technique \cite{shi2017pdd}, which enables updating the optimization variables in parallel and thus can potentially reduce the computational time substantially as compared to the algorithms in \cite{Wu2018_journal, wu2019beamforming} if a multi-core processor with parallel computing capability is available.\footnote{Note that although we consider the same multiuser MISO system in this paper as that in \cite{wu2019beamforming}, the investigated optimization problems, transmission protocols, proposed algorithms and considered channel models in these two works are all different.} Numerical results show that the proposed PDD-based algorithm can achieve near-optimal performance. Furthermore, it is found that as the channel deterministic components become dominant and/or the channel correlation is high in the considered channel model, the rate loss of the proposed TTS optimization with S-CSI as compared to that assuming ideal I-CSI is greatly reduced.

Next, we consider the general multiuser case. Different from the single-user case, deriving closed-form expressions of the achievable average rates of all users in terms of the IRS phase shifts only is difficult because we are unable to obtain the optimal transmit precoding vectors as explicit functions of the IRS phase shifts. To make the problem tractable, we propose an iterative TTS stochastic successive convex approximation (SSCA) algorithm, where in each iteration, a quadratic surrogate of the objective function is constructed based on some appropriately generated channel realizations/samples and the current phase shifts. Then, by employing the Lagrange dual method to solve the resultant quadratic optimization problem, the phase shifts are iteratively  updated with low complexity. On the other hand, with fixed IRS phase shifts, the short-term transmit precoding optimization problems over different channel realizations are efficiently  solved by applying the weighted minimum mean-squared error (WMMSE) algorithm \cite{Shi2011WMMSE}. Numerical results validate the effectiveness of the proposed algorithm and show that using IRS with practical discrete phase shifts under S-CSI can still improve the rate performance significantly over the conventional system without IRS. Moreover, we draw useful insights into the effects of the channel deterministic components and correlation on the proposed TTS algorithm performance.

To the best of our knowledge, this is the first work on the TTS beamforming optimization for IRS-aided communication systems and the new contributions of this paper in view of the existing literature are summarized as follows:
	
1) A new TTS joint active and passive beamforming scheme for an IRS-aided multiuser MISO system is proposed to reduce the channel training overhead and passive beamforming design complexity, where the long-term discrete IRS phase shifts are optimized based on S-CSI and the short-term transmit precoding vectors at the AP are designed according to the effective I-CSI.

2)  To solve the considered TTS optimization problem efficiently, a new PDD-based algorithm and a new SSCA algorithm are respectively proposed for the single-user and multiuser cases. Both algorithms constitute efficient variable updating steps, which either admit closed-form solutions or can be carried out via simple iterative procedures.

3) Extensive numerical results are presented to validate the effectiveness of the proposed TTS transmission protocol and algorithms. The impacts of the IRS channel deterministic Rician components and correlation coefficients on the system performance are investigated and useful insights are drawn.

The rest of the paper is organized as follows. In Section \ref{section_system_model}, we present the system model, the proposed transmission protocol and the corresponding TTS problem formulation. In Sections \ref{section_single_user} and \ref{section_multiuser}, we propose efficient algorithms to solve the TTS problems in the single-user and multiuser cases, respectively. In Section \ref{Section_simulation}, numerical results are provided to evaluate the performance of the proposed algorithms.  Finally, we conclude the paper in Section \ref{section_conclusion}. 

\emph{Notations}: Scalars, vectors and matrices are respectively denoted by lower/upper case, boldface lower case and boldface upper case letters. For an arbitrary matrix $\mathbf{A}$, $\mathbf{A}^T$, $\mathbf{A}^*$ and $\mathbf{A}^{H}$ denote its transpose, conjugate and conjugate transpose, respectively, and $\mathbf{A}^{-1}$ denotes the inverse of a square matrix $\mathbf{A}$. $\textrm{sum}(\mathbf{A})$ denotes the summation of all elements in $\mathbf{A}$ and $\mathbf{A}(m,n)$ denotes the element on the $m$-th row and $n$-column of matrix $\mathbf{A}$. $\|\cdot\|$ and $\|\cdot\|_{\infty}$ denote the Euclidean norm and infinity norm of a complex vector, respectively, and $|\cdot|$ denotes the absolute value of a complex scalar or the cardinality of a finite set. $\odot$ and $\otimes$ denote the Hadamard product and the Kronecker product. For any numbers $x_1,\cdots,x_N$, $\textrm{diag}(x_1,\cdots,x_N)$ denotes a diagonal matrix with $x_1,\cdots,x_N$ being its diagonal elements and $\textrm{diag}(\mathbf{A})$ denotes a vector which contains the diagonal elements of matrix $\mathbf{A}$. The letter $j$ will be used to represent $\sqrt{-1}$ when there is no ambiguity. For a complex number $x$, $\Re \{x\}$ denotes its real part. $\mathbb{C}^{n\times m}$ denotes the space of $n\times m$ complex matrices. $\mathbf{I}$ and $\mathbf{1}$ denote an identity matrix and an all-one matrix/vector with appropriate dimensions, respectively. $\mathbb{E}\{\cdot\}$ represents the statistical expectation operator. $\lfloor x \rfloor$ denotes the maximum integer no larger than $x$. The set difference is defined as $\mathcal{A}\backslash \mathcal{B} \triangleq \{x| x\in\mathcal{A},x\notin \mathcal{B}\}$.

\section{System Model and Problem Formulation} \label{section_system_model}
\subsection{System Model}
\begin{figure}[t] 
	\centering
	\scalebox{0.4}{\includegraphics{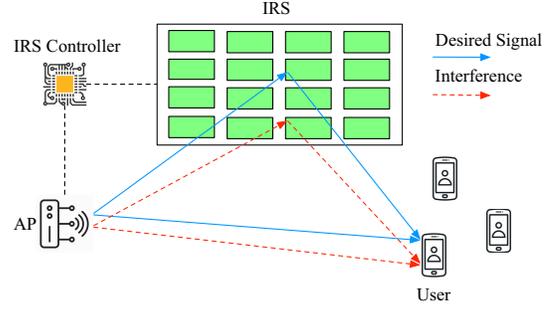}}
	\caption{An IRS-aided multiuser MISO downlink system.}
	\label{fig:system_model}
\end{figure}

As shown in Fig. \ref{fig:system_model}, we consider a multiuser MISO downlink communication system where an IRS equipped with $N$ reflecting elements is deployed to enhance the communications from an AP with $M$ antennas to $K$ single-antenna users. The users are assumed to be in the vicinity of the IRS and of low mobility. The IRS is attached to a smart controller that is able to communicate with the AP via a separate backhaul link for coordinating transmission and exchanging information, such as CSI and IRS phase shifts \cite{Wu2019Magazine}. Since the signal transmitted through the AP-IRS-user link suffers from the double path loss, the signals reflected by IRS two or more times are ignored \cite{Wu2019Magazine, Ozdogan2020WCL}.

Let $\mathbf{G} \in \mathbb{C}^{N\times M}$, $\mathbf{h}_{r,k} \in \mathbb{C}^{N\times 1}$ and $\mathbf{h}_{d,k} \in \mathbb{C}^{M\times 1}$ denote the baseband equivalent channels of the AP-IRS, IRS-user $k$ ($k \in \mathcal{K}\triangleq\{1,\cdots, K\}$) and AP-user $k$ links, respectively. Then, the received signal of user $k$ can be expressed as
\begin{equation} \label{receive_signal}
y_k = (\mathbf{h}_{r,k}^H \mathbf{\bm{\Theta}} \mathbf{G}+\mathbf{h}_{d,k}^H) \mathbf{x} + n_k,
\end{equation}
where $\bm{\Theta}$ denotes an $N \times N$ diagonal reflection coefficient matrix (also known as the passive beamforming matrix \cite{Wu2018_journal}), which can be written as $\bm{\Theta} = \textrm{diag}(\phi_1,\phi_2,\cdots,\phi_N )$, $\phi_n= a_n e^{j\theta_n}$, $a_n\in [0,1], \theta_n \in [0,2\pi), \forall n \in \mathcal{N} \triangleq\{1,\cdots,N\}$; and $n_k$ denotes the independent and identically distributed (i.i.d.) complex additive white Gaussian noise (AWGN) at the receiver of user $k$ with zero mean and variance $\sigma_k^2$. In addition, $\mathbf{x}$ denotes the complex baseband signal transmitted by the AP and can be written as
\begin{equation} \label{transmit_signal}
\mathbf{x} = \sum\limits_{k\in\mathcal{K}}\mathbf{w}_k s_k,
\end{equation}
where $s_k$ represents the information symbol for user $k$ and $s_k$'s are modeled as i.i.d. circularly symmetric complex Gaussian (CSCG) random variables with zero mean and unit variance; $\mathbf{w}_k \in \mathbb{C}^{M \times 1}$ denotes the transmit precoding vector for user $k$. Note that for ease of implementation and cost reduction, we restrict the phase shift at each reflecting element of the IRS to a set of discrete values. Besides, although the reflection amplitude $\{a_n\}$ can be adjusted in the interval $[0,1]$ theoretically\cite{Wu2019Magazine}, it is practically costly to control $\{a_n\}$ and $\{\phi_n\}$ independently and simultaneously. Therefore, we assume $a_n=1,\forall n \in\mathcal{N}$ in this paper to maximize the signal reflection of the IRS \cite{Nayeri2018, Wu2018_journal,wu2019beamforming}. Let $Q$ denote the number of control bits for phase-shifting per IRS element and by assuming that the discrete phase-shift values are obtained by uniformly quantizing the interval $[0,2\pi)$, we have 
\begin{equation} \label{discrete_constraints}
\begin{array}{l}
 \phi_n \in \mathcal{F} \triangleq \Big\{\phi_n| \phi_n = a_n e^{j\theta_n}, \\
\quad\quad \; \theta_n \in \{0,\frac{2\pi}{L},\cdots, \frac{2\pi(L-1)}{L}\}, a_n = 1\Big\},\;\forall n \in \mathcal{N},
 \end{array}
\end{equation}
where $L=2^Q$. Ideally, when $Q \rightarrow \infty$, each element can have any phase-shift value within $[0, 2\pi)$, which is referred to as the continuous phase-shift case. 

Based on \eqref{receive_signal} and \eqref{transmit_signal}, the received signal-to-interference-plus-noise ratio (SINR) of user $k$ can be expressed as
\begin{equation} \label{SINR}
\textrm{SINR}_k = \frac{|(\mathbf{h}_{r,k}^H \bm{\Theta} \mathbf{G}+ \mathbf{h}_{d,k}^H) \mathbf{w}_k|^2}{\sum\limits_{j \in \mathcal{K}\backslash k}|(\mathbf{h}_{r,k}^H \bm{\Theta} \mathbf{G}+ \mathbf{h}_{d,k}^H) \mathbf{w}_j|^2 + \sigma_k^2},
\end{equation}
and the corresponding achievable rate in bits/second/Hertz (bits/s/Hz) is $r_k = \log_2(1+\textrm{SINR}_k)$.

\subsection{Channel Model}
Since both line-of-sight (LoS) and non-LoS (NLoS)  components may exist in practical channels and due to the insufficient angular spread of the scattering environment and/or closely spaced antennas/reflecting elements, we model the AP-user, AP-IRS and IRS-user channels as the general spatially correlated Rician fading channels \cite{McKay2005}. Moreover, since the distances between the IRS and its served users are relatively small, IRS elements reflect signals with a finite angular spread and a user-location dependent mean angle in practice \cite{Yin2013}. Therefore, we assume that the channel statistics of the IRS-user links are user-location dependent. In contrast, since the distances between the AP and users are much larger than those between the IRS and users, we assume that the second-order statistics of the AP-user links are identical for all users. Specifically, the channel of the IRS-user $k$ link can be modeled as
\begin{equation}
\mathbf{h}_{r,k} = \sqrt{\frac{\beta_{Iu}}{1+\beta_{Iu}}}\bar{\mathbf{z}}_{r,k} + \sqrt{\frac{1}{1+\beta_{Iu}}} \bm{\Phi}_{r,k}^{1/2} \mathbf{z}_{r,k} ,
\end{equation}
where $\mathbf{z}_{r,k} \in \mathbb{C}^{N \times 1}$ has i.i.d. CSCG entries with zero mean and unit variance accounting for small-scale fading (assumed to be Rayleigh fading); $\bm{\Phi}_{r,k} \in \mathbb{C}^{N\times N}$ is the spatial correlation matrix between the IRS and user $k$; $\bar{\mathbf{z}}_{r,k}$ and $\beta_{Iu}$ denote the deterministic component and the Rician factor, respectively. Similarly, for AP-IRS and AP-user links, we have \cite{McKay2005}
\begin{equation}
\begin{aligned}
\mathbf{G} = \sqrt{\frac{\beta_{AI}}{1+\beta_{AI}}} \bar{\mathbf{F}} + \sqrt{\frac{1}{1+\beta_{AI}}} \bm{\Phi}_r^{1/2} \mathbf{F} \bm{\Phi}_d^{1/2} ,\\
\mathbf{h}_{d,k} = \sqrt{\frac{\beta_{Au}}{1+\beta_{Au}}} \bar{\mathbf{z}}_{d,k} + \sqrt{\frac{1}{1+\beta_{Au}}} \bm{\Phi}_d^{1/2}\mathbf{z}_{d,k},
\end{aligned}
\end{equation}
where $\bar{\mathbf{F}}$ and $ \bar{\mathbf{z}}_{d,k}$ denote the deterministic components, $\mathbf{F} \in \mathbb{C}^{N\times M}$ and $\mathbf{z}_{d,k} \in \mathbb{C}^{M\times 1 }$ denote the corresponding small-scale fading components similar as $\mathbf{z}_{r,k}$, $\bm{\Phi}_d$ and $\bm{\Phi}_r$ denote the AP transmit correlation matrix and the IRS receive correlation matrix, respectively. As we consider low-mobility users, the spatial correlation matrices $\tilde{\bm{\Phi}} \triangleq \{ \bm{\Phi}_r,\bm{\Phi}_{r,k}, \bm{\Phi}_d\}$ (assumed to be real matrices) and deterministic components $\mathbf{H}_{\textrm{LoS}} \triangleq\{\bar{\mathbf{z}}_{r,k}, \bar{\mathbf{z}}_{d,k}, \bar{\mathbf{F}}\}$, i.e., the S-CSI, may change slowly in practice. In contrast, the I-CSI can vary much more rapidly due to the phase variations induced by the relatively slight movement of the users and/or the scattering objects in the environment. Note that there are two types of ``correlation'' in the considered channel model. One is due to the existence of deterministic components while the other is due to the scattering environment and the antenna/reflecting element configurations.\footnote{Note that these two types of ``correlation'' may have a similar effect on the CSI. For example, for the AP-IRS channel matrix $\mathbf{G}$, when $\beta_{AI}\rightarrow \infty$ (i.e., deterministic component dominates the channel), it follows that $\mathbf{G} \approx \bar{\mathbf{F}}$, which is usually a rank-one matrix under the far-field array model. In contrast, when $\beta_{AI} = 0$ but the channels tend to be fully correlated, i.e., $\bm{\Phi}_r \rightarrow \mathbf{1}$ and $\bm{\Phi}_d \rightarrow \mathbf{1}$, it follows that $\mathbf{G} = \frac{1}{\sqrt{NM}}\textrm{sum}(\mathbf{F})\mathbf{1}$, which is also a rank-one matrix.} For simplicity, we integrate the terms involving the Rician factors into $\{\bar{\mathbf{z}}_{r,k}, {\mathbf{z}}_{r,k}, \bar{\mathbf{F}}, {\mathbf{F}}, \bar{\mathbf{z}}_{d,k}, {\mathbf{z}}_{d,k}\}$, then the channel models can be equivalently and more concisely rewritten as
$\mathbf{h}_{r,k} = \bar{\mathbf{z}}_{r,k} +  \bm{\Phi}_{r,k}^{1/2} \mathbf{z}_{r,k},\;
\mathbf{G} =  \bar{\mathbf{F}} + \bm{\Phi}_r^{1/2} \mathbf{F} \bm{\Phi}_d^{1/2}$, and $
\mathbf{h}_{d,k} =  \bar{\mathbf{z}}_{d,k} + \bm{\Phi}_d^{1/2}\mathbf{z}_{d,k}$.

\subsection{Transmission Protocol}
\begin{figure}[!hhh]  
	\centering
	\scalebox{0.4}{\includegraphics{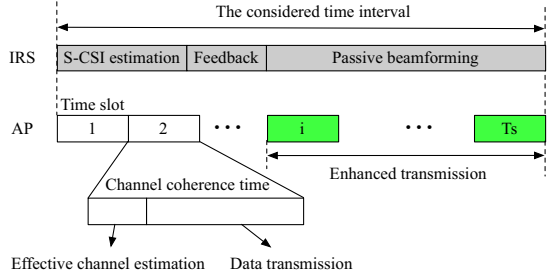}}
	\caption{Frame structure of the proposed transmission protocol.} 
	\label{fig:figure1} 
\end{figure}

Since the acquisition of the effective fading channels $\{\mathbf{h}_k \triangleq  \mathbf{G}^H \bm{\Theta}^H \mathbf{h}_{r,k} + \mathbf{h}_{d,k}\}$ from the AP to users with given fixed IRS phase shifts is much easier in practice as compared with that of the IRS-associated channels $\mathbf{G}$ and $\{\mathbf{h}_{r,k}\}$, we propose a hierarchical transmission protocol in this paper. Specifically, we focus on a time interval within which the S-CSI of all links is assumed to remain constant, as shown in Fig. \ref{fig:figure1}. The considered time interval consists of $T_s \gg 1$ time slots and can be divided into three transmission phases. The small-scale fading coefficients $\{\mathbf{z}_{r,k}\}$, $\{\mathbf{z}_{d,k}\}$ and $\mathbf{F}$ are assumed to be constant within each time slot (or equivalently, the I-CSI $\tilde{\mathbf{H}} \triangleq \{\mathbf{h}_{d,k},\mathbf{h}_{r,k}, \mathbf{G}\}$). Each time slot is further divided into two sub-slots where the first sub-slot is for effective fading channel estimation and the second is for data transmission.

In the first phase, the IRS is in the sensing mode and the channel statistical information between the IRS and the AP/users can be estimated by resorting to the dedicated sensors/receiving circuits at the IRS and leveraging the pilots and/or data transmitted in both uplink and downlink using standard mean and covariance matrices estimation techniques \cite{Mestre2008, Werner2008}.\footnote{How to efficiently deploy the sensors on IRS and estimate the required S-CSI based on their low-resolution sensing measurements is an interesting problem that is worth further investigating in future work.}  Note that in this phase, the AP serves the users by only utilizing the direct channels $\{\mathbf{h}_{d,k} \}$ estimated in the first sub-slot of each time slot as if the IRS does not exist. The direct channels $\{\mathbf{h}_{d,k} \}$ in this phase are equivalent to the effective fading channels $\{\mathbf{h}_k\}$ since the IRS is in the sensing mode. In the second phase, based on the measured S-CSI of the AP-IRS-user links (fed back by the IRS) and that of the AP-user links (measured in Phase I), the AP computes the IRS passive beamforming matrix $\bm{\Theta}$, and sends it to the IRS through the dedicated backhaul link. Finally, in the third phase, the IRS is switched to the reflection mode with the phase shifts given in $\bm{\Theta}$ to enhance the transmissions from the AP to the users. Specifically, for each time slot during this phase, the AP estimates the effective I-CSI $\{{\mathbf{h}}_k\}$ by applying the channel estimation methods in traditional MIMO systems \cite{MISO_channel_estimation} and designs its transmit precoding vectors $\{\mathbf{w}_k\}$ accordingly.\footnote{In the sequel of this paper, to focus on the TTS beamforming optimization, we assume for simplicity that the above S-CSI and effective I-CSI are perfectly known at the AP at the beginning of the considered time interval and each of its time slots, respectively. Further investigation into the TTS design under imperfect CSI is left for our future work.}

Note that since the number of reflecting elements at the IRS, $N$, is usually much larger than that of transmit antennas at the AP, the effective CSI
$\{{\mathbf{h}}_k\}$ usually has a much smaller dimension than the full channel ensemble $\tilde{\mathbf{H}}$. Therefore, compared to the existing transmit and reflect beamforming optimization in e.g., \cite{Wu2018_journal, wu2019beamforming, guo2019weighted, zhang2019capacity}, based on the I-CSI of all channels, the beamforming design complexity and channel estimation overhead can be significantly reduced by the proposed new protocol based on S-CSI. Besides, for fast-varying channels, using I-CSI may not be helpful as previously acquired I-CSI will become outdated quickly, which renders the proposed S-CSI-based protocol more suitable.

\subsection{Problem Formulation}
In this paper, we aim to maximize the average weighted sum-rate of all the users by jointly optimizing the short-term active transmit precoding at the AP and long-term passive reflect beamforming at the IRS, subject to the maximum transmit power constraint at the AP. The corresponding optimization problem can be formulated as
\begin{equation} \label{TTS_problem_detail}
\begin{aligned}
\max\limits_{\bm{\Theta}} \; & \mathbb{E}\left\{ \max\limits_{\{\mathbf{w}_k \}}\sum\limits_{k\in\mathcal{K}}\alpha_k\log_2\left(1+\textrm{SINR}_k\right) \right\}\\
\textrm{s.t.} \; & \sum\limits_{k \in \mathcal{K}} \|\mathbf{w}_k\|^2 \leq P,\;
\phi_n \in \mathcal{F},\;\forall n \in \mathcal{N},
\end{aligned}
\end{equation}
where $\alpha_k$ represents the weight/priority of user $k$ and $P$ denotes the total transmit power budget. The inner rate-maximization problem in \eqref{TTS_problem_detail} is over the short-term transmit precoding in each time slot/channel realization for given phase shifts $\bm{\Theta}$ at the IRS, while the outer rate-maximization problem is over the long-term IRS phase shifts, where the expectation is taken over all channels' random realizations within the considered time interval.

Furthermore, let $\Omega \triangleq \{\mathbf{w}_k(\tilde{\mathbf{H}}) \in \mathcal{X},\;\forall \tilde{\mathbf{H}}\}$ denote the set of transmit precoding vectors (each as a function of the random instantaneous channel $\tilde{\mathbf{H}}$) that satisfy the constraint $\sum\nolimits_{k \in \mathcal{K}} \|\mathbf{w}_k\|^2 \leq P$, and define $\bar{\mathbf{r}}(\bm{\Theta},\Omega) = [\bar{r}_1(\bm{\Theta},\Omega) ,\cdots,\bar{r}_K(\bm{\Theta},\Omega) ]^T$ as the achievable average rate vector, where $\bar{r}_k(\bm{\Theta},\Omega) = \mathbb{E}\{r_k(\bm{\Theta},\{\mathbf{w}_k(\tilde{\mathbf{H}})\}, \tilde{\mathbf{H}})\}$.\footnote{For notation simplicity, we may drop the arguments in $r_k(\bm{\Theta},\{\mathbf{w}_k(\tilde{\mathbf{H}})\}, \tilde{\mathbf{H}})$ and $\bar{\mathbf{r}}(\bm{\Theta},\Omega)$, and simply use $r_k$ and $\bar{\mathbf{r}}$ respectively in the sequel of this paper when there is no ambiguity.} Then, problem \eqref{TTS_problem_detail} can be rewritten in a more compact form as
\begin{equation} \label{TTS_problem}
\max\limits_{\{\mathbf{w}_k\},\;\textrm{diag}(\bm{\Theta}) \in \mathcal{F}^N} \; \bm{\alpha}^T\bar{\mathbf{r}}(\bm{\Theta},\Omega),
\end{equation}
where $\bm{\alpha} \triangleq [\alpha_1,\cdots, \alpha_K]^T$ and $\mathcal{F}^N$ is defined as the Cartesian product of $N$ identical sets each given by $\mathcal{F}$.
Problem \eqref{TTS_problem} is challenging to solve because 1) the short-term transmit precoding vectors $\{\mathbf{w}_k\}$ and
the long-term IRS phase shifts $\bm{\Theta}$ are intricately coupled in the objective function; 2) a closed-form expression of the achievable average rate of each user, $\bar{r}_k(\bm{\Theta},\Omega)$, for given either $\bm{\Theta}$ or $\Omega$, is difficult to obtain in general; and 3) it is  a mixed-integer non-linear program (MINLP) even for $K=1$. Generally, there is no efficient method for solving the non-convex problem \eqref{TTS_problem} optimally. In the next two sections, we propose two efficient algorithms to solve problem \eqref{TTS_problem} sub-optimally in the single-user and multiuser cases, respectively.

\newtheorem{prop}{Proposition}

\section{Single-User Case} \label{section_single_user}
In this section, we consider the single-user case, i.e., $K = 1$, where there is no multiuser interference. Accordingly, problem \eqref{TTS_problem} reduces to (by dropping the user index)
\begin{equation}\label{TTS_problem_single_user}
	\max\limits_{\mathbf{w},\; \textrm{diag}(\bm{\Theta}) \in \mathcal{F}^N} \; \bar{r}(\bm{\Theta},\Omega).
\end{equation}
Note that in \cite{Han2019}, a similar problem to \eqref{TTS_problem_single_user} was considered for the single-user case by exploiting the S-CSI, while in this section we address problem \eqref{TTS_problem_single_user} under our considered fading channel model which is more general than that in \cite{Han2019}. For any given phase-shift matrix $\bm{\Theta}$, it is well-known that the maximum-ratio transmission (MRT) at the AP is optimal \cite{Tse2005, wu2019beamforming}, i.e.,  
\begin{equation} \label{optimal_w_single_user}
\mathbf{w}^{\textrm{opt}} = \sqrt{P}\frac{(\mathbf{h}_r^H \bm{\Theta} \mathbf{G} + \mathbf{h}_d^H)^H}{\|\mathbf{h}_r^H \bm{\Theta} \mathbf{G} + \mathbf{h}_d^H\|}.
\end{equation}
Based on \eqref{optimal_w_single_user}, we obtain the following proposition. 
\begin{prop} \label{prop_single_user}
\emph{In the single-user case, the achievable rate $\bar{r}(\bm{\Theta},\Omega)$ in \eqref{TTS_problem_single_user} is upper-bounded by $\log_2\big(1+(\mathbf{v}^H \bm{\Phi} \mathbf{v} + \mathbf{v}^H \mathbf{b} +  \mathbf{b}^H \mathbf{v}+\bar{\mathbf{z}}_d^H \bar{\mathbf{z}}_d+\frac{1}{1+\beta_{Au}} \sum\nolimits_{i=1}^M \bm{\Phi}_d(i,i))/\sigma^2 \big)$, where $\mathbf{v} = [v_1,\cdots,v_N]^T = {\textrm{diag}}(\bm{\Theta}^*) \in \mathbb{C}^{N\times 1}$, $\mathbf{b} = {\textrm{diag}}\{\bar{\mathbf{z}}_r^H\}\bar{\mathbf{F}}  \bar{\mathbf{z}}_d$, 
\begin{equation}
	\begin{aligned}
	\bm{\Phi}  & =  {\textrm{diag}}\{\bar{\mathbf{z}}_r^H\} \bar{\mathbf{F}} \bar{\mathbf{F}}^H  {\textrm{diag}}\{\bar{\mathbf{z}}_r\} \\
	& \quad + \frac{1}{1+\beta_{AI}}\left( \sum\limits_{i=1}^M\lambda_i \right) {\textrm{diag}}\{ \bar{\mathbf{z}}_r^H\} \bm{\Phi}_r {\textrm{diag}}\{\bar{\mathbf{z}}_r\} \\
	& \quad + \frac{1}{1+\beta_{Iu}}  (\bm{\Phi}_{r,u}  \odot (\bar{\mathbf{F}} \bar{\mathbf{F}}^H)) \\
	& \quad + \left( \sum\limits_{i=1}^M\lambda_i \right)  \frac{1}{1+\beta_{AI}} \frac{1}{1+\beta_{Iu}}( \bm{\Phi}_{r,u} \odot \bm{\Phi}_r),
	\end{aligned}
\end{equation}
$\lambda_i$ denotes the $i$-th eigenvalue of $\bm{\Phi}_d$, and $\bm{\Phi}_{r,u}$ denotes the correlation matrix between the IRS and the user.
}
\end{prop}
\newtheorem{remark}{Remark}
\begin{proof}
	Please refer to Appendix \ref{appendix_proof_proposition1}.
\end{proof}
\noindent Based on Proposition \ref{prop_single_user}, we can remove the $\log_2(\cdot)$ operator in the logarithmic rate upper-bound function due to its monotonicity and ignore the constant terms, then problem \eqref{TTS_problem_single_user} can be approximated by the following deterministic problem:
\begin{equation} \label{statistic_problem}
\begin{aligned}
\max\limits_{\mathbf{v}} \; & \mathbf{v}^H \bm{\Phi} \mathbf{v} + \mathbf{v}^H \mathbf{b} +  \mathbf{b}^H \mathbf{v}\\
{\textrm{s.t.}}\;& v_n \in \mathcal{F},\;\forall n \in \mathcal{N}.
\end{aligned}
\end{equation}
For problem \eqref{statistic_problem}, it can be shown that when $\beta_{Au}=\beta_{AI}=\beta_{Iu}\rightarrow \infty$, the objective function is reduced to $\mathbf{v}^H {\textrm{diag}}\{\bar{\mathbf{z}}_r^H\} \bar{\mathbf{F}} \bar{\mathbf{F}}^H  {\textrm{diag}}\{\bar{\mathbf{z}}_r\} \mathbf{v} + \mathbf{v}^H \mathbf{b} +  \mathbf{b}^H\mathbf{v}$ and in this case, the IRS phase-shift vector $\mathbf{v}$ is optimized based on the deterministic component $\mathbf{H}_{\textrm{LoS}}$ only, i.e., $\{\bar{\mathbf{z}}_{r}, \bar{\mathbf{z}}_{d}, \bar{\mathbf{F}}\}$. In contrast, when $\beta_{AI}=\beta_{Iu}\rightarrow 0$, i.e., in the case of NLoS environment, the optimal solution to \eqref{statistic_problem} is shown in Appendix \ref{appendix_proof_proposition2} to be $\bm{\mathbf{v}} = \bar{\phi} \mathbf{1}$, for any phase shift $|\bar{\phi}| = 1$, i.e., the phase shifts of all elements at the IRS should be identical. In general, $\mathbf{v}$ should be properly designed to strike a balance between the deterministic and NLoS channels.

Although problem \eqref{statistic_problem} is much simplified compared to problem \eqref{TTS_problem_single_user}, it is still a non-convex quadratic programming  problem with discrete constraints that is NP-hard in general. In \cite{Wu2018_journal} and \cite{wu2019beamforming}, various methods were proposed to address a similar problem by using e.g.,  the SDR method and the BCD method. However, the SDR method incurs a high complexity and the BCD method requires to update the phase shifts one-by-one iteratively. To reduce their computational time, we propose a new and alternative algorithm in this paper, namely, the PDD-based algorithm, which enables the optimization of IRS phase shifts in parallel. Besides, the proposed PDD-based algorithm is able to handle the discrete IRS phase-shift constraints and each of its iterations can be executed in closed-form, as shown next.

\subsection{Proposed PDD-based Algorithm} \label{SU_PDD_section}
To facilitate the optimization of $\mathbf{v}$ in parallel, we introduce an auxiliary variable $\mathbf{u} = [u_1,u_2,\cdots, u_N]^T \in \mathbb{C}^{N\times 1}$ ($u_n=c_n e^{j \vartheta_n}$), which satisfies $\mathbf{u} = \mathbf{v}$. As a result, problem \eqref{statistic_problem} is equivalent to
\begin{equation}  \label{statistic_proble_AADP_eq}
\begin{aligned}
\max\limits_{\mathbf{v},\; \mathbf{u}} \; & \mathbf{v}^H \bm{\Phi} \mathbf{v} + \mathbf{v}^H \mathbf{b} +  \mathbf{b}^H \mathbf{v} \\
\textrm{s.t.}\;& \mathbf{v} = \mathbf{u}, \;u_n \in \mathcal{F},\;\forall n \in\mathcal{N}.
\end{aligned}
\end{equation}
To address problem \eqref{statistic_proble_AADP_eq}, we propose the PDD-based algorithm consisting of two loops. In the outer loop, we update the dual variable associated with the constraint $\mathbf{v} = \mathbf{u}$ and the penalty parameter (as will be introduced later in this subsection), while in the inner loop, we apply the block successive upper-bound minimization (BSUM) method \cite{Hong2016} to iteratively optimize the primal variables in different blocks. Specifically, we can write the augmented Lagrangian (AL) problem of \eqref{statistic_proble_AADP_eq} as follows \cite{Zhao2019CL, Zhao2019TPS}:
\begin{equation} \label{AL_problem}
\begin{aligned}
	\min\limits_{\mathbf{v},\;\mathbf{u}} \; & -\mathbf{v}^H \bm{\Phi}  \mathbf{v}  - \mathbf{v}^H \mathbf{b} -  \mathbf{b}^H \mathbf{v} + \frac{1}{2\rho}  \|\mathbf{v}-\mathbf{u}+\rho \bm{\lambda}\|^2\\
	\textrm{s.t.}\; &  u_n \in \mathcal{F},\;\forall n \in\mathcal{N},\; \|\mathbf{v}\|^2 \leq N,
\end{aligned} 
\end{equation}
where $\rho$ is the penalty parameter and $\bm{\lambda} = [\lambda_1,\cdots,\lambda_N]^T$ denotes the dual variable vector associated with the constraint $ \mathbf{v} = \mathbf{u}$. Note that $\|\mathbf{v}\|^2 \leq N$ is a new constraint that is added without loss of optimality due to $|v_n|\leq 1$ and the necessity of this constraint will be clarified later. Then, we partition all the optimization variables in \eqref{AL_problem} into two blocks, i.e., $\mathbf{v}$ and $\mathbf{u}$, and optimize them iteratively in the inner loop as follows.

\textbf{Step 1} (optimizing $\mathbf{v}$ for given $\mathbf{u}$): this subproblem is given by
		\begin{equation} \label{v_subproblem_ori}
	\begin{aligned}
	\min\limits_{\mathbf{v}}\; & \frac{1}{2\rho}\|\mathbf{v} - \mathbf{u}+\rho \bm{\lambda}\|^2-\mathbf{v}^H \bm{\Phi}  \mathbf{v}  - \mathbf{v}^H \mathbf{b} -  \mathbf{b}^H \mathbf{v} \\
	\textrm{s.t.}\; & \|\mathbf{v}\|^2 \leq N.
	\end{aligned}
	\end{equation}
Since its constraint is convex and the objective function can be expressed as a difference of two convex functions when $\mathbf{u}$ is fixed, we can apply the BSUM method to solve it approximately. Specifically, by resorting to the first-order Taylor expansion at a given point $\bar{\mathbf{v}}$, problem \eqref{v_subproblem_ori} can be approximated by
	\begin{equation} \label{v_subproblem}
	\begin{aligned}
	\min\limits_{\mathbf{v}}\; & \frac{1}{2\rho}\|\mathbf{v} - \mathbf{u}+\rho \bm{\lambda}\|^2 - 2\Re\{(\bm{\Phi} \bar{\mathbf{v}})^H(\mathbf{v}-\bar{\mathbf{v}})\}  - \mathbf{v}^H \mathbf{b} -  \mathbf{b}^H \mathbf{v} \\
	\textrm{s.t.}\; & \|\mathbf{v}\|^2 \leq N.
	\end{aligned}
	\end{equation} 
It is not difficult to observe that problem \eqref{v_subproblem} is now a convex quadratically constrained quadratic program (QCQP) with only one constraint whose optimal solution can be obtained by exploiting the first-order optimality condition as follows:
\begin{equation} \label{PDD_update_1}
\mathbf{v}^{\textrm{opt}} = \left\{ \begin{array}{l}
\mathbf{c}, \;\textrm{if}\; \|\mathbf{c}\|^2 \leq N,\\
\frac{\mathbf{c}}{\sqrt{\|\mathbf{c}\|^2/N}},\;\textrm{otherwise},
\end{array}
\right.
\end{equation}
where $ \mathbf{c}= \mathbf{u}-\rho \bm{\lambda} + 2\rho \bm{\Phi}  \bar{\mathbf{v}} + 2\rho\mathbf{b}$. Note that if the constraint $\|\mathbf{v}\|^2 \leq N$ is absent, one has to set $\rho \leq \frac{1}{2\lambda_{\textrm{max}}(\bm{\Phi})}$ in order to make problem \eqref{v_subproblem_ori} bounded. Therefore, if $\lambda_{\textrm{max}}(\bm{\Phi})$ is large, then the initial penalty (i.e., $\frac{1}{2\rho}$) is large, which will severely restrict the search space of the proposed algorithm. Thus, the constraint $\|\mathbf{v}\|^2 \leq N$ is necessary. 

\textbf{Step 2} (optimizing $\mathbf{u}$ for given $\mathbf{v}$): this subproblem can be written as (ignoring constant terms)
\begin{equation} \label{u_subproblem}
\min\limits_{\mathbf{u}} \;  \|\mathbf{v}-\mathbf{u}+\rho \bm{\lambda}\|^2\quad 
\textrm{s.t.}\;  u_n \in \mathcal{F},\;\forall n \in\mathcal{N}.
\end{equation}
Due to the fact that $\{u_n\}$ are decoupled in both the objective function and the constraints of problem \eqref{u_subproblem}, we can easily obtain the optimal phase shifts of this subproblem in parallel as follows:  first obtain the optimal continuous phase-shift solution as $\vartheta_n= \angle (v_n+\rho \lambda_n)$, and then map $\vartheta_n$ to the nearest discrete value in $ \{0,\frac{2\pi}{L},\cdots, \frac{2\pi(L-1)}{L}\}$.  

Next, we consider the outer loop, where the dual variable $\bm{\lambda}$ can be updated by
\begin{equation} \label{dual_update}
\bm{\lambda} = \bm{\lambda} + \frac{1}{\rho}\left(\mathbf{v}-\mathbf{u}\right),
\end{equation}
which is a dual ascend step.
The main steps of the proposed PDD-based algorithm are summarized in Algorithm \ref{PDD_algorithm_adj}, where $c < 1$ is a constant scaling factor that is used to increase the value of the penalty term involved in problem \eqref{AL_problem} in each outer iteration. We note that the penalty parameter $\rho$ can affect the convergence of the
PDD-based algorithm. Specifically, if $\rho$ decreases too fast, the AL problem \eqref{AL_problem} will become ill-conditioned and it may lead to undesired results or stuck to some unfavorable points; on the other hand, if $\rho$ decreases too slow, it may affect the convergence speed of the PDD-based algorithm. Therefore, the parameter $c$ should be appropriately chosen to control the decreasing speed of $\rho$.\footnote{In our simulations, we find that choosing an arbitrary value of $c$ from the interval $[0.7,0.99]$ will not lead to significant performance variations, which means that the proposed PDD-based algorithm is quite robust under different values of $c$.} Besides, according to \cite{shi2017pdd}, Algorithm \ref{PDD_algorithm_adj} is guaranteed to converge to a set of stationary solutions of problem \eqref{statistic_problem} in the continuous phase-shift case. While for the discrete phase-shift case, the solution obtained by solving problem \eqref{AL_problem} always satisfies the equality constraint $\mathbf{v} = \mathbf{u}$, as $\rho \rightarrow 0$ ($\frac{1}{\rho} \rightarrow \infty$) \cite{Bertsekas1999}. Therefore, Algorithm \ref{PDD_algorithm_adj} is able to converge to a high-quality suboptimal solution, as will be verified later in Section \ref{Section_simulation}.

\begin{algorithm}[!h] \small
	\caption{{Proposed PDD-based Algorithm for Solving Problem \eqref{TTS_problem_single_user}}} \label{PDD_algorithm_adj}
	\begin{algorithmic}[1]
		\STATE Initialize $\mathbf{v}^0, \mathbf{u}^0$ and $c$, set the outer iteration number $i_{\textrm{out}} \leftarrow 0$.
		\REPEAT
		\STATE Set the inner iteration number $i_{\textrm{in}} \leftarrow 0$.
		\REPEAT
		\STATE Update ${\mathbf{v}}$ and $\mathbf{u}$ according to Steps 1-2.
		\STATE Update $i_{\textrm{in}} \leftarrow i_{\textrm{in}} + 1$.
		\UNTIL{The fractional decrease of the objective value of \eqref{AL_problem} is below a certain threshold $\epsilon_{\textrm{in}} > 0$ or the maximum number of inner iterations is reached.}
		\STATE Update the dual variable by \eqref{dual_update} and decrease the penalty parameter as $\rho  \leftarrow c\rho $.
		\STATE Update $i_{\textrm{out}} \leftarrow i_{\textrm{out}} + 1$.
		\UNTIL{The constraint violation $\|\mathbf{v}-\mathbf{u}\|_{\infty}$ is below a certain threshold $\epsilon_{\textrm{out}} > 0$. }
	\end{algorithmic}
\end{algorithm}

\subsection{Complexity Analysis}
The complexity of Algorithm \ref{PDD_algorithm_adj} is mainly due to solving problem \eqref{v_subproblem}, which can be shown of $\mathcal{O}(N^2)$. Thus, the overall complexity of Algorithm \ref{PDD_algorithm_adj} is $\mathcal{O}(I_oI_iN^2)$, where $I_o$ and $I_i$ denote the maximum outer and inner iteration numbers. In contrast, the worst-case complexity of the SDR method in \cite{Wu2018_journal} is $\mathcal{O}(N^{6.5})$ and that of the successive refinement algorithm in \cite{wu2019beamforming} is $\mathcal{O}(IN^2)$, where $I$ denotes the number of iterations required for convergence. To summarize, the complexity of the proposed Algorithm \ref{PDD_algorithm_adj} is much lower than the SDR method. Besides, although Algorithm \ref{PDD_algorithm_adj} and the successive refinement algorithm exhibit the same complexity order, using Algorithm \ref{PDD_algorithm_adj} can reduce the computational time, especially for practically large IRS, if a multi-core processor is available (see the parallel update in Step 2 of Section \ref{SU_PDD_section}).

\section{Multiuser Case} \label{section_multiuser}
In this section, we address the multiuser case where multiple users are assumed to share the same time-frequency resource and the multiuser interference exists in general. Specifically, we leverage the stochastic optimization framework in \cite{Liu2018TOSCA} to propose a novel SSCA algorithm, where the phase shifts at the IRS (i.e., the long-term variables) are updated by solving the outer rate-maximization problem in \eqref{TTS_problem_detail} with randomly generated channel samples, and the transmit precoding vectors at the AP (i.e., the short-term variables) are optimized in each time slot by applying the WMMSE method \cite{Shi2011WMMSE}.

\subsection{Short-Term Optimization Problem} \label{section_short_term}
At each time slot $m \in [1, T_s]$, the AP first acquires the effective fading channel $\mathbf{H}(m) \triangleq \{\mathbf{h}_1(m),\cdots,\mathbf{h}_K(m) \}$ with fixed phase shifts $\mathbf{v}$. Then, the AP designs the short-term transmit precoding vectors $\{\mathbf{w}_k\}$, by applying the WMMSE method to solve the following problem,
\begin{equation} \label{short_term_problem}
\begin{aligned}
\max\limits_{\{\mathbf{w}_k\}} \; &  \sum\limits_{k \in \mathcal{K}} \alpha_k \log_2\left(1+\frac{|\mathbf{h}_k^H(m) \mathbf{w}_k|^2}{\sum\limits_{j \in\mathcal{K}\backslash k}|\mathbf{h}_k^H(m) \mathbf{w}_j|^2 + \sigma_k^2}\right) \\ 
\textrm{s.t.} \; & \sum\limits_{k \in \mathcal{K}} \|\mathbf{w}_k\|^2 \leq P,
\end{aligned}
\end{equation}
for given $\mathbf{H}(m)$. Note that $\{\mathbf{w}_k\}$ are optimized based on the effective fading channels $\{\mathbf{h}_k\}$ only. The basic idea of the WMMSE method is to first transform problem \eqref{short_term_problem} into an equivalent WMMSE optimization problem, and then update the optimization variables alternately until convergence is achieved. The details of this method can be found in \cite{Shi2011WMMSE} where it shows that a stationary solution of problem \eqref{short_term_problem} can be obtained; thus, they are omitted for brevity.

\subsection{Long-Term Optimization Problem} \label{Section_long_term}
When the S-CSI is obtained, the AP optimizes the IRS phase shifts $\mathbf{v}$ by solving problem \eqref{TTS_problem}. Note that unlike the single-use case for which the closed-form MRT-based optimal transmit precoding is available, the optimized precoding vectors via WMMSE in the multiuser case cannot be expressed explicitly, thus it is difficult to obtain the closed-form expression of $ \bm{\alpha}^T\bar{\mathbf{r}}$ (as well as its lower or upper bounds) in terms of $\mathbf{v}$. To address this issue, we propose an efficient algorithm, where $\mathbf{v}$ is updated iteratively by maximizing a concave surrogate function of $ \bm{\alpha}^T\bar{\mathbf{r}}$, denoted by $ \bar{f}^t(\mathbf{v})$, with $t$ denoting the iteration index. Furthermore, we relax the amplitudes of $\mathbf{v}$ to be in the interval $[0,1]$, which will be shown to help accelerate the convergence of the proposed algorithm by simulation in Section \ref{Section_simulation}. Note that we can simply set $v_n \leftarrow e^{j \angle{v_n}},\forall n \in \mathcal{N}$ to recover the unit-modulus solution of $\mathbf{v}$ after the convergence is reached. Let $\mathbf{v}^{t-1}$ denote the IRS phase-shift vector obtained from the $(t-1)$-th iteration. Then the $t$-th iteration of the proposed algorithm, for any $t \geq 1$, is described as follows.

First, $T_H$ new channel samples $\{\tilde{\mathbf{H}}^t(l)\}_{l=\{1,\cdots,T_H\}} \triangleq \{\mathbf{h}_{r,k}(l),\mathbf{h}_{d,k}(l), \mathbf{G}(l)\}_{l=\{1,\cdots,T_H\}}^t$ are randomly generated according to the S-CSI $\tilde{\bm{\Phi}}$ and $\mathbf{H}_{\textrm{LoS}}$. Based on them, we update the surrogate function to obtain $\bar{f}^t(\mathbf{v}) $, which can be viewed as a concave approximation of the objective function $\bm{\alpha}^T\bar{\mathbf{r}}$ of problem \eqref{TTS_problem}.
Specifically, based on $\{\tilde{\mathbf{H}}^t(l)\}_{l=\{1,\cdots,T_H\}}$ and the phase-shift vector $\mathbf{v}^{t-1}$, $ \bar{f}^t(\mathbf{v})$ is obtained as
\begin{equation} \label{surrogate_function}
\bar{f}^t(\mathbf{v}) = \bm{\alpha}^T \hat{\mathbf{r}}^t + 2\Re\{(\mathbf{f}^t)^H(\mathbf{v}-\mathbf{v}^{t-1})\} -\tau \|\mathbf{v}-\mathbf{v}^{t-1}\|^2,
\end{equation}
where the last term is added to ensure that $-\bar{f}^t(\mathbf{v})$ is uniformly and strongly convex with respect to (w.r.t.) $\mathbf{v}$ so as to guarantee the convergence of the proposed algorithm \cite{Liu2018TOSCA} with any constant $ \tau> 0$; $\hat{\mathbf{r}}^t = [\hat{r}_1^t,\cdots,\hat{r}_K^t]^T$ is an approximation of the achievable average rate vector, which is updated as
\begin{equation} \label{approximate_average_rate}
\hat{r}_k^t = (1-\rho_t)\hat{r}_k^{t-1} +\rho_t \sum\limits_{l=1}^{T_H}  \frac{\alpha_k r_k(\mathbf{v}^{t-1},\{\mathbf{w}^t_k(l)\};\tilde{\mathbf{H}}^t(l))}{T_H},
\end{equation}
with $\hat{r}_k^{0} = 0,\;\forall k \in \mathcal{K}$, $\rho_t$ satisfies Assumption 1 (i.e., Assumption 5 in \cite{Liu2018TOSCA}), which will be specified later, $\mathbf{w}^t_k(l)$ denotes the transmit precoding vector corresponds to the $l$-th generated channel sample with fixed $\mathbf{v}^{t-1}$, i.e., $\mathbf{w}^t_k(l) \triangleq \mathbf{w}_k(\mathbf{v}^{t-1}, \tilde{\mathbf{H}}^t(l))$, and $\mathbf{f}^t = [f_1^t,\cdots,f_N^t]^T$ is an approximation of the partial derivative $\nabla_{\mathbf{v}^*} \bm{\alpha}^T\bar{\mathbf{r}}$, which can be similarly updated as
\begin{equation} \label{approximate_derivative}
\begin{aligned}
\mathbf{F}^t & =  (1-\rho_t) \mathbf{F}^{t-1} +  \rho_t \sum\limits_{l=1}^{T_H} \frac{ \mathbf{J}_{\mathbf{v}^*}(\mathbf{v}^{t-1},\{\mathbf{w}^t_k(l)\}; \tilde{\mathbf{H}}^t(l)) }{T_H},\\
\mathbf{f}^t & =  \mathbf{F}^t \nabla_{\bar{\mathbf{r}}}\hat{\mathbf{r}}^t = \mathbf{F}^t [\alpha_1,\cdots,\alpha_K]^T,
\end{aligned}
\end{equation}
where $\mathbf{F}^{0} = \mathbf{0}$, $\mathbf{J}_{\mathbf{v}^*}(\mathbf{v}^{t-1},\{\mathbf{w}^t_k(l)\}; \tilde{\mathbf{H}}^t(l))$ is the Jacobian matrix of the achievable rate vector $\mathbf{r}(\mathbf{v},\{\mathbf{w}_k\};\tilde{\mathbf{H}}) \triangleq [r_1(\mathbf{v},\{\mathbf{w}_k\};\tilde{\mathbf{H}}),\cdots, r_K(\mathbf{v},\{\mathbf{w}_k\};\tilde{\mathbf{H}})]^T$
w.r.t. $\mathbf{v}^*$ and its detailed expression is given in Appendix \ref{Jacobian_matrix}. Note that $\mathbf{F}^t$ is an approximation of $\mathbb{E}\{\mathbf{J}_{\mathbf{v}^*}(\mathbf{v}^{t-1},\mathbf{w}_k(\mathbf{v}^t,\tilde{\mathbf{H}}); \tilde{\mathbf{H}})\}$. The iterative approximations $\bar{f}^t(\mathbf{v})$ and $\mathbf{f}^t$ can converge to the true values of the objective function $ \bm{\alpha}^T\bar{\mathbf{r}}$ of problem \eqref{TTS_problem} and its gradient w.r.t. $\mathbf{v}^*$, as $t \rightarrow \infty$\cite[Theorem 1]{Liu2018TOSCA}.
Therefore, based on the randomly generated channel samples $\{\tilde{\mathbf{H}}^t(l)\}$ at the beginning of each iteration and the corresponding solutions $\{\mathbf{w}^t_k(l)\}$ of the short-term problems, the achievable average rate $\bar{\mathbf{r}}(\mathbf{v}, \{\mathbf{w}_k\}; \tilde{\mathbf{H}})$, although not expressed explicitly, can be approximated by updating $ \bar{\mathbf{r}}$ and $\nabla_{\mathbf{v}^*} \bm{\alpha}^T \bar{\mathbf{r}}$ in an iterative manner as in \eqref{approximate_average_rate} and \eqref{approximate_derivative}.
With \eqref{surrogate_function}, we only need to solve the following quadratic optimization problem, 
\begin{equation} \label{passive_problem}
\begin{aligned}
\max\limits_{\mathbf{v}} \; & \bar{f}^t(\mathbf{v})\\ 
\textrm{s.t.}\; & |v_n| \leq 1,\;\forall n \in \mathcal{N},
\end{aligned}
\end{equation}
which is convex and its optimal solution can be obtained in closed-form as shown next. 
\begin{remark}
\emph{Note that to make the overall problem tractable, we have ignored the discrete constraints and relaxed $\{v_n\}$ as continuous variables in problem \eqref{passive_problem}. After obtaining the optimized IRS phase shifts $\mathbf{v}$, we project each of its entries independently onto $\mathcal{F}$ to obtain a unit-modulus solution, i.e.,
\begin{equation}
\hat{v}_n = \arg\min\limits_{\hat{v}_n \in \mathcal{F}}\; |\angle v_n - \angle \hat{v}_n|, \;\forall n \in \mathcal{N}.
\end{equation} 
It was shown in \cite{wu2019beamforming} that using $Q=2$ or $3$ bits is practically sufficient to achieve near-optimal performance. In this paper, although we consider a different problem, it will be shown in Section \ref{Section_simulation} that when $Q \geq 2 $ bits, the performance loss due to discrete phase shifts is also negligible.}
\end{remark}  
Note that in problem \eqref{passive_problem}, all the optimization variables $\{v_n\}$ can be fully decoupled and thus we can optimize each of them independently in parallel. As such, problem \eqref{passive_problem} w.r.t. $v_n$ can be equivalently rewritten as (by ignoring constant terms)
\begin{equation} \label{long_term_problem}
\begin{aligned}
\min\limits_{v_n}\;	& \tau v_n^* v_n - \tau v_nv_n^{(t-1)*} - \tau v_n^* v_n^{t-1} -  f_n^{t*}v_n - v_n^*f_n^{t} \\
\textrm{s.t.}\; & v_n^* v_n \leq 1,
\end{aligned}
\end{equation}
which is a convex optimization problem. By resorting to the Lagrange duality method, we can obtain the Lagrangian dual function as $\mathcal{L}(v_n,\lambda) = \tau v_n^* v_n - \tau v_n v_n^{(t-1)*} - \tau v_n^* v_n^{t-1} - f_n^{t*}v_n - v_n^*f_n^{t} + \lambda (v_n^* v_n- 1)$, 
where $\lambda$ is the dual variable associated with the constraint in \eqref{long_term_problem}. Then, the optimal solution of problem \eqref{long_term_problem} can be obtained in closed-form as follows: if $|v_n^{t-1} + \frac{f_n^t}{\tau}|\leq 1$, we have $\bar{v}_n^t = v_n^{t-1}+ \frac{f_n^t}{\tau}$; otherwise, we have $\bar{v}_n^{t} = \frac{\tau v_n^{t-1} + f_n^t}{\tau+\lambda^{\textrm{opt}}}$, where $\lambda^{\textrm{opt}} $ denotes the optimal dual variable and is given by $\lambda^{\textrm{opt}}  = |\tau v_n^{t-1} + f_n^t| - \tau$.
Therefore, the long-term IRS phase shifts $\mathbf{v}$ can be updated according to
\begin{equation} \label{long_term_update}
\mathbf{v}^{t} = (1-\gamma_t) \mathbf{v}^{t-1} + \gamma_t \bar{\mathbf{v}}^t,
\end{equation}
where $\gamma_t$ is an iteration-dependent constant that satisfies the following assumption (referred to as Assumption 1): 
1) $\rho_t \rightarrow 0$, $\frac{1}{\rho_t} \leq \mathcal{O}(t^\kappa)$ for some $\kappa \in (0,1)$, $\sum\nolimits_{t} (\rho_t)^2 < \infty$;
2) $\gamma_t \rightarrow 0$, $\sum\nolimits_{t} \gamma_t= \infty$, $\sum\nolimits_{t}(\gamma_t)^2 \leq \infty$; and 
3) $\lim_{t\rightarrow \infty} \frac{\gamma_t}{\rho_t} = 0$.
The above procedure is repeated until convergence and the overall algorithm is summarized in Algorithm \ref{SSCA_algorithm}.

The convergence of Algorithm \ref{SSCA_algorithm} is analyzed as follows. Consider problem \eqref{TTS_problem_detail} with continuous phase shifts and adjustable amplitudes between $[0,1]$ and refer to it as problem $\mathcal{C}$. If problem $\mathcal{C}$ has at least one stationary solution, then every
limit point $\mathbf{v}^{\textrm{lim}}$ of the sequence $\{\mathbf{v}^t\}_{t=1}^{\infty}$ generated by Algorithm \ref{SSCA_algorithm} is a stationary point of the long-term (outer rate-maximization) problem of $\mathcal{C}$ when the transmit precoding vectors are obtained by the WMMSE method. Besides, Algorithm \ref{SSCA_algorithm} almost surely converges to the set of stationary solutions of problem $\mathcal{C}$ and the detailed proof can be found in \cite{Liu2018TOSCA, Liu2019CRAN}. Besides, the advantages of the proposed Algorithm \ref{SSCA_algorithm} are summarized as follows: 1) by iteratively constructing a surrogate function based on randomly generated channel samples and the corresponding solutions of the short-term optimization problems, it is able to resolve the difficulty caused by the unavailability of the closed-form expression of $ \bm{\alpha}^T\bar{\mathbf{r}}$, 2) the long-term IRS phase shifts $\mathbf{v}$ and short-term transmit precoding vectors $\{\mathbf{w}_k\}$ are \emph{jointly} optimized to maximize $ \bm{\alpha}^T\bar{\mathbf{r}}$, and 3) the optimization can be conducted through a sequence of simple and efficient updates on the variables.

\begin{algorithm}[!h]\small
	\caption{{Proposed TTS SSCA Algorithm for Solving Problem \eqref{TTS_problem}}} \label{SSCA_algorithm}
	\begin{algorithmic}  
		\STATE \textbf{Input}: $\{\rho_t\}$, $\{\gamma_t\}$ and $T_H$. \textbf{Initialize}: $\mathbf{v}^0$ and $t=1$. 
		\STATE \textbf{Step 1}: (Long-term optimization with given S-CSI):
		\STATE \begin{itemize}
			\item Generate $T_H$ new channel samples according to the known S-CSI $\tilde{\bm{\Phi}}$ and $\mathbf{H}_{\textrm{LoS}}$.
		 \item Update the surrogate function by \eqref{surrogate_function}, where $\{\mathbf{w}^t_k(l)\}$ are obtained by applying the WMMSE method to solve problem \eqref{short_term_problem} with given generated channel samples and fixed $\mathbf{v}^{t-1}$.
		 \item Solve problem \eqref{long_term_problem} to obtain the optimal $\bar{\mathbf{v}}^t$ and update $\mathbf{v}^{t}$ according to \eqref{long_term_update}.
		 \item Let $t = t + 1$ and return to \textbf{Step 1}. Repeat the above until convergence. Denote the converged phase-shift vector as $\mathbf{v}$. 
		\end{itemize}
	    \STATE \textbf{Step 2}: (Short-term optimization at each time slot $m \in [1,T_s]$):
		\STATE \begin{itemize}
			\item Apply the WMMSE method with given $\mathbf{v}$ and $\mathbf{H}(m)$ to obtain the short-term variables $\{\mathbf{w}_k\}$.
		\end{itemize}
	\end{algorithmic}
\end{algorithm}

\subsection{Complexity Analysis}
From the above, it is observed that the complexity of Algorithm \ref{SSCA_algorithm} is mainly due to computing $\{\mathbf{w}_k(l)\}_{l=\{1,\cdots,T_H\}}$ for the generated channel samples in the long-term optimization problem. For each $l \in \{1,\cdots,T_H\}$, the WMMSE method is applied to obtain the corresponding transmit precoding vectors, whose complexity is dominated by the matrix inversion operation required for updating $ \{\mathbf{w}_k \}$, which is $\mathcal{O}(JKM^3)$, where $J$ denotes the number of WMMSE iterations. Accordingly, the complexity for updating the long-term IRS phase-shift vector $\mathbf{v}$ is $\mathcal{O}(I (T_H JKM^3+KNM))$, where $I$ denotes the iteration number required for the phase-shift optimization in Section \ref{Section_long_term}. Therefore, the overall complexity of Algorithm \ref{SSCA_algorithm} is given by $\mathcal{O}(I (T_H JKM^3+KNM) + T_sJKM^3)$.

\section{Simulation Results} \label{Section_simulation}
In this section, we provide numerical results to evaluate the performance of the proposed algorithms and draw useful insights. The distance-dependent path loss is modeled as $L = C_0\left(\frac{d_{\textrm{link}}}{D_0}\right)^{-\alpha}$, where $C_0$ is the path loss at the reference distance $D_0 = 1$ meter (m), $d_{\textrm{link}}$ represents the individual link distance and $\alpha$ denotes the path loss exponent. The path loss exponents of the AP-user, AP-IRS and IRS-user links are denoted by $\alpha_{Au}$, $\alpha_{AI}$ and $\alpha_{Iu}$, respectively. We assume that the IRS is deployed to serve the users that suffer from severe signal attenuation in the AP-user direct link and thus we set $\alpha_{Au} = 3.4$, $\alpha_{AI} = 2.2$ and $\alpha_{Iu} = 3$, i.e., the path loss exponent of the AP-user link is larger than those of the AP-IRS and IRS-user links. In our simulations, a three-dimensional coordinate system is considered where the AP (equipped with a uniform linear array (ULA)) and the IRS (equipped with a uniform rectangular array (UPA)) are located on the $x$-axis and $y$-$z$ plane (or parallel to the $x$-$z$ plane), respectively. In the single-user case, we set $N = N_yN_z$ where $N_y$ and $N_z$ denote the numbers of reflecting elements along the $y$-axis and $z$-axis, respectively, while in the multiuser case, we set $N = N_xN_z$ with $N_x$ denoting the number of reflecting elements along the $x$-axis. For the purpose of exposition, we fix $N_y = 4$ in the single-user case and $N_x=4$ in the multiuser case. The reference antenna/element at the AP/IRS are located at $(d_v, 0, 0)$ and $(0, d_0=50\;\textrm{m}, 3\;\textrm{m})$. Moreover, we consider the following exponential correlation model for $\bm{\Phi}_{d}$ \cite{Loyka2001,Xia2015,Yi2019}:
\begin{equation} \label{expo_model}
\bm{\Phi}_d(i,j) = \left\{ 
\begin{array}{l}
r_d^{j-i},\; \textrm{if}\; i \leq j,\\
\bm{\Phi}_d(j,i),\;\textrm{if}\; i>j,
\end{array}
\right.
\end{equation}
where $0\leq r_d\leq 1$ is the correlation coefficient. $\bm{\Phi}_r$ is modeled as $\bm{\Phi}_r = \bm{\Phi}_{r}^h \otimes \bm{\Phi}_{r}^v$ \cite{Choi2014}, where $\bm{\Phi}_{r}^h$ and $\bm{\Phi}_{r}^v$ denote the spatial correlation matrices of the horizontal and vertical domains, respectively, and are similarly defined as in \eqref{expo_model} with $r_r$ denoting the correlation coefficient. $\{\bm{\Phi}_{r,k} =\bm{\Phi}_{r,k}^h \otimes \bm{\Phi}_{r,k}^v \}$ ($\bm{\Phi}_{r,u}$ for the single-user case) are similarly modeled as $\bm{\Phi}_r$  with $\{r_{r,k}\}$ ($r_{r,u}$ for the single-user case) denoting the corresponding correlation coefficients. The deterministic component of each channel is modeled as a random matrix/vector with i.i.d. CSCG entries of zero mean and unit variance, and kept fixed during the entire time interval. Other system parameters are set as follows unless otherwise specified: $\sigma_k^2 = -80$ dBm, $P = 5$ dBm, $C_0 = -30$ dB, $N=40$, and for the single-user case, we set $M=4$, $\epsilon_{\textrm{in}}=10^{-4}$, $\epsilon_{\textrm{out}}=10^{-6}$, $ \beta_{AI} = \beta_{Iu} = 3$ dB and $\beta_{Au}  = -3$ dB, while for the multiuser case, we let $M=6$, $K=4$, $T_H = 10$, $T_s = 2000$, $\alpha_k = 1,\forall k \in \mathcal{K}$, $c=0.95$, $\tau = 0.01$, $\rho_t = t^{-0.8}$, $\gamma_t = t^{-1}$,  $\beta_{AI} = \beta_{Iu}   = 5$ dB and $ \beta_{Au} = -5$ dB. All the results are averaged over $2000$ independent channel realizations.

\subsection{Single-User Case} \label{section_singleuser}
\begin{figure}[!hhh]  
	\centering
	\scalebox{0.4}{\includegraphics{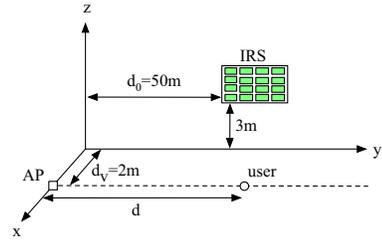}}
	\caption{Simulation setup of the single-user case.}
	\label{fig:SU_setup}
\end{figure}

We first consider the single-user case where the user is assumed to move along the line $(2\;\textrm{m},d,0)$, as shown in Fig. \ref{fig:SU_setup}.
For comparison, we adopt the following five benchmark schemes: 1) the SDR method with Gaussian randomization \cite{Wu2018_journal}, 2) a naive scheme where the phase shifts at the IRS are obtained by Algorithm \ref{PDD_algorithm_adj} with I-CSI at the first time slot and then kept fixed for all the subsequent time slots,  3) a single-timescale scheme where both $\mathbf{v}$ and $\mathbf{w}$ are optimized based on the S-CSI and kept fixed for all the time slots, 4) the random phase-shift scheme where the phase shifts at the IRS are randomly generated at each time slot, and 5) the conventional scheme by using the MRT beamforming at the AP, but without the IRS.

\begin{figure}[!hhh]
	\centering
	\scalebox{0.35}{\includegraphics{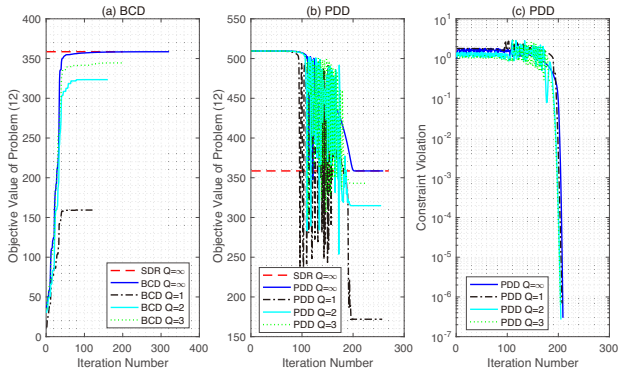}}
	\caption{Convergence behaviors of the BCD algorithm versus the proposed PDD-based algorithm.}
	\label{fig:SU_comvergence} 
\end{figure}

First, we compare in Fig. \ref{fig:SU_comvergence} the convergence behaviors of the BCD algorithm \cite{wu2019beamforming} and the proposed PDD-based algorithm (Algorithm \ref{PDD_algorithm_adj}) with $d=50$ m.
In our simulations, using the SDR method with Gaussian randomization to solve problem \eqref{statistic_problem} can achieve near-optimal performance, as shown for a similar problem in \cite{Wu2018_journal}. Therefore, its performance is considered as an upper bound for the BCD and the PDD-based algorithms. From Fig. \ref{fig:SU_comvergence} (a) and (b), it is observed that the BCD algorithm is monotonically convergent, while this is not the case for the PDD-based algorithm in general. Furthermore, there exist fluctuations of the objective value in the initial few iterations of the PDD-based algorithm. This is mainly because when the initial penalty is relatively small, the solutions obtained by the PDD-based algorithm do not satisfy $v_n=u_n,\;\forall n \in\mathcal{N}$, thus resulting in the oscillatory behavior. As the penalty increases with the iteration number, the constraint violation, i.e., $\|\mathbf{v}-\mathbf{u}\|_{\infty}$, is forced to approach the predefined accuracy $\epsilon_{\textrm{out}}$, as shown in Fig. \ref{fig:SU_comvergence} (c). As a result, the PDD-based algorithm is guaranteed to converge which can be observed from Fig. \ref{fig:SU_comvergence} (b). Moreover, one can observe from Fig. \ref{fig:SU_comvergence} (a) and (b) that the PDD-based algorithm can achieve near-optimal performance in the continuous phase-shift case, i.e., $Q=\infty$, and for the discrete phase-shift case (e.g. $Q=1$), its performance is similar to that of the BCD algorithm.

\subsubsection{Impact of the AP-user distance $d$} 
\begin{figure}[!hhh]
	\centering
	\scalebox{0.35}{\includegraphics{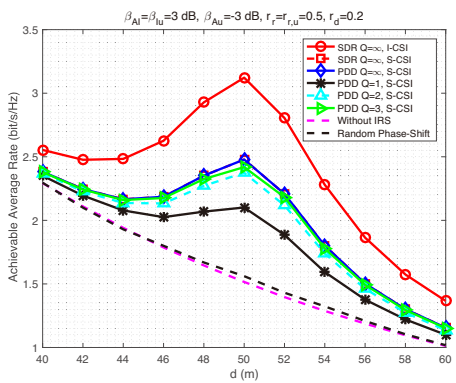}}
	\caption{Achievable average rate versus the AP-user distance $d$ in the single-user case.}
	\label{fig:SU_compare_distance}
\end{figure}

In Fig. \ref{fig:SU_compare_distance}, we plot the achievable average rate of the user versus the AP-user distance $d$ with $r_r=r_{r,u} = 0.5$ and $r_d=0.2$. It is observed that when the user lies in the neighborhood of the IRS, the achievable average rate by using 1-bit phase shifters ($Q=1$) with S-CSI is significantly higher than that without IRS and that with random phase shift at the IRS. This means that IRS is practically useful by creating a ``signal hot spot'' even with coarse and low-cost phase shifters and S-CSI. Moreover, it is observed that using IRS with 1-bit phase shifters results in a considerable performance loss as compared to the ideal case with continuous phase shifters. However, this performance gap can be effectively reduced by using higher-resolution phase shifters, e.g., $Q=2$ and $Q=3$.

\subsubsection{Impact of the Rician factor} 
\begin{figure}[!hhh]
	\centering
	\scalebox{0.35}{\includegraphics{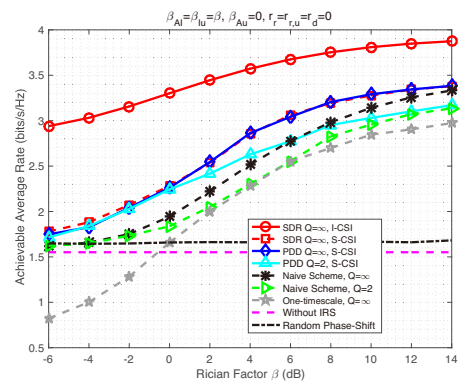}}
	\caption{Achievable average rate versus the Rician factor in the single-user case.}
	\label{fig:SU_compare_K_factor}
\end{figure}

In Fig. \ref{fig:SU_compare_K_factor}, we plot the achievable average rate versus the Rician factor by fixing $d=50$ m. To focus on the effect of Rician factor on the system rate performance, we assume $\beta_{AI}  = \beta_{Iu}  = \beta$, $\beta_{Au} = 0$ and $r_r = r_{r,u}=r_d=0$, i.e., the AP-user link is assumed to follow Rayleigh fading (no deterministic components exist due to blockage) while the AP-IRS and IRS-user links are assumed to follow uncorrelated Rician fading. It is observed from Fig. \ref{fig:SU_compare_K_factor} that the performance of all algorithms with both S-CSI and I-CSI improves with $\beta$. This is expected since as $\beta$ increases, the AP-IRS channel becomes more correlated which is highly beneficial for achieving the maximum beamforming gain in the single-user case. In particular, for the S-CSI case, another important reason is that when $\beta$ increases, the AP-IRS-user link becomes more deterministic, thus rendering the proposed scheme based on S-CSI to be more effective. Furthermore, we can observe that the performance gap between the two cases (I-CSI and S-CSI) eventually approaches a constant when $\beta$ is sufficiently large. This is because the AP-user link is assumed to be Rayleigh fading, thus for the S-CSI case, no statistical information can be extracted and exploited to further improve the achievable average rate. It is also observed that the PDD-based algorithm outperforms the naive and the single-timescale schemes since they do not fully exploit the S-CSI and I-CSI, respectively.

\subsubsection{Impact of the correlation coefficients $r_r$ and $r_{r,u}$} 
\begin{figure}[!hhh]
	\centering
	\scalebox{0.35}{\includegraphics{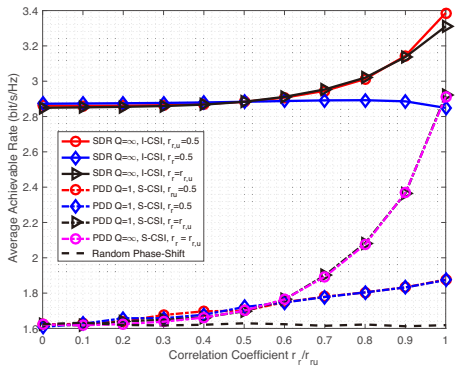}}
	\caption{Achievable average rate versus the IRS correlation coefficients $r_r$ and $r_{r,u}$ with fixed $r_d = 0$ in the single-user case.}
	\label{fig:SU_compare_correlation}
\end{figure}

In Fig. \ref{fig:SU_compare_correlation}, we investigate the achievable average rate versus the correlation coefficients $r_r$ and $r_{r,u}$. For ease of comparison, we set $d=50$ m, $r_d = 0$ and $\beta_{Au}=\beta_{AI} = \beta_{Iu}  = 0$. From Fig. \ref{fig:SU_compare_correlation}, we observe that for the I-CSI case, the achievable average rate improves with the increasing of $r_r$ (with the reason similar to that in Fig. \ref{fig:SU_compare_K_factor}), but this does not hold when increasing $r_{r,u}$. To be specific, the performance with $\{r_r=0.5, r_{r,u}=1\}$/$\{r_r=1, r_{r,u}=1\}$ is inferior to that with $\{r_r=0.5, r_{r,u}=0.9\}$/$\{r_r=1, r_{r,u}=0.5\}$. This is because when $r_{r,u} $ is close to $1$, i.e., the IRS-user channel is fully correlated (i.e., the entries in $\mathbf{h}_r$ are almost identical), the degree of freedom (DoF) when adjusting the transmit precoding vector for signal alignment at the user becomes very limited. Furthermore, it is observed that the benefit brought by increasing $r_r$ and $r_{r,u}$ is more pronounced for the algorithms based on S-CSI than that based on I-CSI. We can also observe that for the S-CSI-based schemes, the effects of increasing $r_r$ or $r_{r,u}$ are similar, and the best performance is achieved when both $r_r$ and $r_{r,u}$ are close to $1$.
Besides, using IRS with discrete phase shifters incurs only negligible rate loss in this case. This is mainly because when $r_r$ is small, exploiting S-CSI only is generally ineffective and thus using continuous phase shifters can only achieve marginal performance gain. When $r_r$ is close to 1, the rate achieved by aligning the phase shifts of all the reflecting elements is already sufficiently high, thus weakening the gain of using continuous phase shifters.

\subsection{Multiuser Case}
\begin{figure}[!hhh]
	\centering
	\scalebox{0.4}{\includegraphics{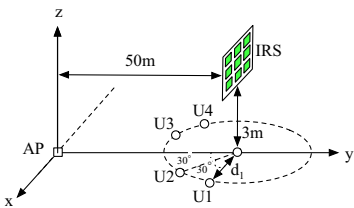}}
	\caption{Simulation setup of the multiuser case.}
	\label{fig:multiuser_setup}
\end{figure}

Next, we consider a multiuser system with four users, denoted by $U_k$'s, $k \in \{1,\cdots,4\}$ and their locations are shown in Fig. \ref{fig:multiuser_setup}, i.e., the users lie on a semicircle centered at $(0,50\;\textrm{m},0)$ with radius $d_1=3$ m. This setup can practically correspond to the case that the IRS is deployed at the cell-edge to cover an area with a high density of users (e.g., a hot-spot scenario). Moreover, to investigate the impacts of the correlation coefficients $\{r_{r,k}\}$ on each user's achievable average rate, we assume $r_{r,k} = \frac{k-1}{3}$, i.e., each IRS-user link has a different correlation level. Similar to the single-user case, four benchmark schemes are considered: 1) an I-CSI-based algorithm which is obtained by combining the WMMSE method and the PDD method, and assuming perfect I-CSI over all time slots, 2) a naive scheme by applying the I-CSI-based algorithm for the first time slot only,  3) the random phase-shift scheme (same as that in the single-user case), and 4) the scheme by applying the WMMSE method in \cite{Shi2011WMMSE}, but without the IRS. We observe by simulation that in the multiuser case, the performance of the single-timescale scheme is even worse than that of the random phase-shift scheme. This is because when the Rician factor of the AP-user links is small, utilizing S-CSI alone results in severe multiuser interference, therefore its performance is not shown here.

\begin{figure}[!hhh]
	\centering
	\scalebox{0.35}{\includegraphics{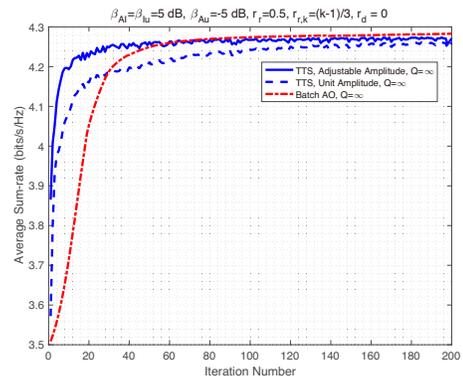}}
	\caption{Convergence behaviors of Algorithm \ref{SSCA_algorithm} and the batch AO algorithm.}
	\label{fig:Convergence_behavior_compare}
\end{figure}

Prior to performance comparison, we first illustrate in Fig. \ref{fig:Convergence_behavior_compare} the convergence behavior of Algorithm 2 by plotting the average sum-rate of the users versus the number of iterations with $r_r=0.5$ and $r_d = 0$. For comparison, we also consider a batch alternating optimization (AO) algorithm \cite{Liu2016BatchAO}. From Fig. \ref{fig:Convergence_behavior_compare}, we can observe that Algorithm \ref{SSCA_algorithm} (with adjustable amplitude versus unit amplitude) and the batch AO algorithm achieve a similar performance when convergence is reached. However, since Algorithm \ref{SSCA_algorithm} with adjustable amplitude converges faster and consumes less storage space as compared with the batch AO algorithm, we only provide the performance of Algorithm \ref{SSCA_algorithm} in the following. Moreover, Algorithm \ref{SSCA_algorithm} with adjustable amplitude also converges faster than that with unit amplitude. This is mainly due to fact that when the amplitudes can be adjusted in the interval $[0,1]$, a larger feasible region can be explored in the first few iterations of the algorithm, which helps accelerate its convergence. 

\begin{figure}[!hhh]
	\centering
	\scalebox{0.35}{\includegraphics{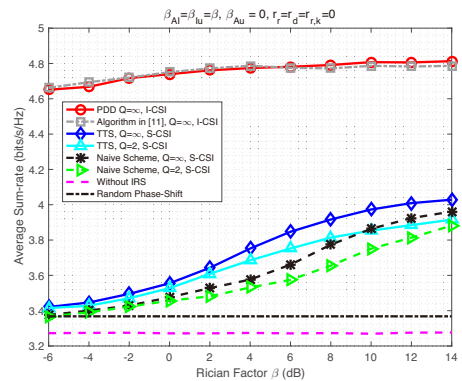}}
	\caption{Average sum-rate versus the Rician factor in the multiuser case.}
	\label{fig:MU_compare_K_factor}
\end{figure}
\subsubsection{Impact of the Rician factor} In Fig. \ref{fig:MU_compare_K_factor}, we investigate the average sum-rate achieved by Algorithm \ref{SSCA_algorithm} versus the Rician factor of the AP-IRS-user links, where we assume $\beta_{AI} = \beta_{Iu}   = \beta$, $\beta_{Au} = 0$, $r_r =r_d=0$ and $r_{r,k} = 0,\;\forall k \in \mathcal{K}$ for simplicity. For the I-CSI-based scheme, we also provide the performance of the algorithm proposed in \cite{guo2019weighted} for comparison. As shown, the PDD-based algorithm achieves a similar performance as the algorithm in \cite{guo2019weighted}. Then, similar to the single-user case shown in Fig. \ref{fig:SU_compare_K_factor}, it can be observed that the performance gap between the schemes based on I-CSI  versus S-CSI decreases with the increasing of $\beta$. The performance gap cannot approach zero under the considered simulation setup, besides the reason mentioned in Fig. \ref{fig:SU_compare_K_factor}, this is also because the multiuser interferences are the performance bottleneck in the multiuser case, therefore if no I-CSI can be exploited in the IRS reflection design to effectively cancel them, the average sum-rate would deteriorate. Furthermore, we observe that different from the proposed algorithm based on S-CSI, the average sum-rate of the random phase-shift scheme and that without IRS are insensitive to the Rician factor.

\begin{figure*}[htbp]
	\centering
	\scalebox{0.4}{\includegraphics{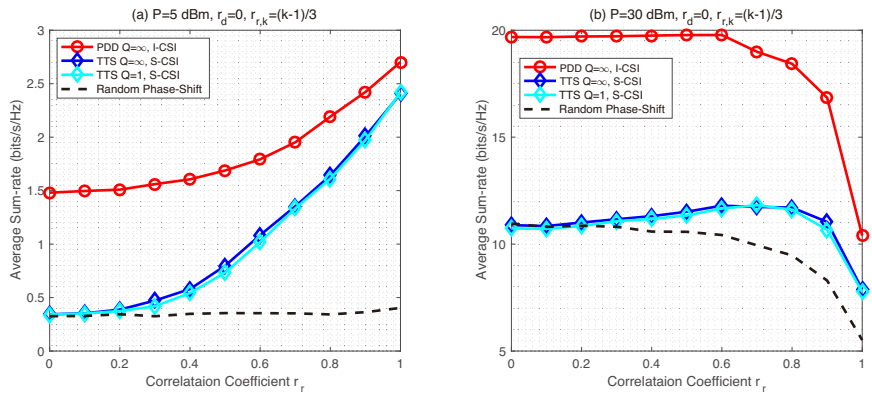}}
	\caption{Average sum-rate versus the IRS correlation coefficients $r_r$ with fixed $r_d = 0$ in the multiuser case.}
	\label{fig:MU_compare_rr}
\end{figure*} 

\subsubsection{Impact of the correlation coefficient $r_r$} In Fig. \ref{fig:MU_compare_rr}, we plot the average sum-rate achieved by Algorithm \ref{SSCA_algorithm} versus the IRS correlation coefficient $r_r$ with fixed $r_d = 0$ and different total transmit power budgets ($P=5$ dBm or $30$ dBm). To focus on studying the effect of $r_r$, the AP-user direct links are assumed to be fully blocked, i.e., $\mathbf{h}_{d,k} = \mathbf{0}, \forall k \in \mathcal{K}$. Note that when $P=5$ dBm, very few users will be scheduled for transmission at each time slot in general according to the solution obtained by solving problem \eqref{TTS_problem}, since the transmit power is limited. In this case, it is observed that the performance of both schemes based on I-CSI or S-CSI improves with the increasing of $r_r$, and the performance gap between them is significantly reduced when $r_r$ is close to $1$. For the I-CSI case, the performance improvement comes from the fact that when $r_r \rightarrow 1$, the AP-IRS channel is nearly rank-one, and thus only one user will be scheduled in each channel realization, which reduces to the single-user case in Section \ref{section_singleuser}. Note that this is quite different from the power minimization problem in \cite{Wu2018_journal} with individual user rate/SINR constraints, for which reducing the rank of $\mathbf{G}$ (or increasing the correlation in $\mathbf{G}$) will decrease the spatial multiplexing gain and result in more severe multiuser interference, thus leading to degraded performance. For the S-CSI case, the average sum-rate increases more rapidly with the increasing of $r_r$, since this reduces the randomness in CSI and enhances the passive beamforming gain. However, when $P=30$ dBm, i.e., at the high SNR region, increasing $r_r$ too much is adverse since in this case the spatial multiplexing gain becomes the performance bottleneck of the system. Besides, similar results can be observed by assuming $r_{r,k}=r_{r,u},\;\forall k \in \mathcal{K}$ and investigating the impact of the correlation coefficient $r_{r,u}$ on the average sum-rate, therefore their details are not shown for brevity.

\subsubsection{Performance comparison with different IRS-user correlation levels} 
\begin{figure*}[htbp]
	\centering
	\scalebox{0.4}{\includegraphics{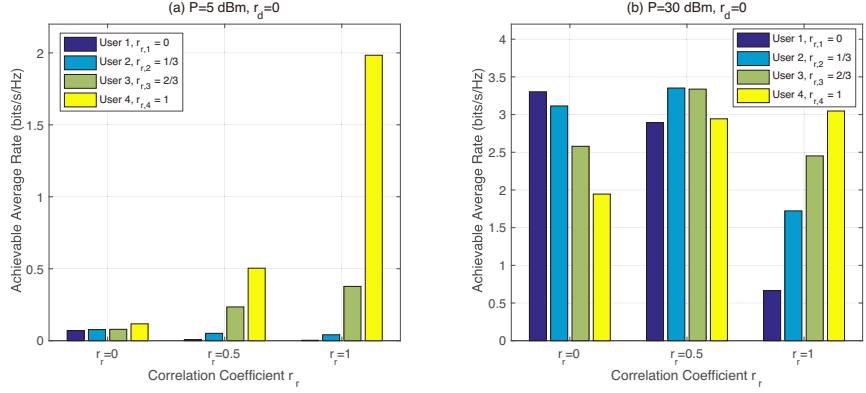}}
	\caption{Achievable average rate of each user with different values of $r_r$ in the multiuser case.}
	\label{fig:MU_compare_user_rate}
\end{figure*}

In Fig. \ref{fig:MU_compare_user_rate}, we investigate the achievable average rate of each user (with different values of $r_{r,k}$) achieved by Algorithm \ref{SSCA_algorithm}, where the simulation parameters are the same as those in Fig. \ref{fig:MU_compare_rr}. As can be seen from Fig. \ref{fig:MU_compare_user_rate} (a), when $P=5$ dBm, the achievable average rate of the user with larger $r_{r,k}$ is always higher than those with smaller ones. This is because when one IRS-user link is more correlated than the others, its S-CSI can be better exploited to improve its achievable rate. Thus, allocating more power to this user is more beneficial for the sum-rate maximization due to the limited total transmit power at the AP. As a result, we can observe that when $r_r=1$, the achievable average rate of user $1$ ($r_{r,1} = 0$) is almost zero, whereas that of user 4 ($r_{r,4} = 1$) is the highest.
Moreover, we note that although larger $r_r$ results in higher sum-rate, this may not be beneficial for achieving the spatial multiplexing gain, since in this case the passive beamforming design favors only a small number of users and user fairness is difficult to guarantee. In Fig. \ref{fig:MU_compare_user_rate} (b), we can observe that the distribution of the achievable rate of each user changes with the IRS receive correlation coefficient $r_r$ when the total transmit power is high. Specifically, if $r_r = 0$, then the AP-IRS channel becomes Rayleigh fading, user $1$ with no channel correlation achieves the best average rate because the channel diversity gain for user $1$ is higher than those of the others when optimizing the transmit precoding vectors. In contrast, when $r_r=1$ (in this case only one user can be supported), the performance of user $4$ is the best since its channel is more deterministic and the optimization of the IRS phase shifts tends to favor this user with more dominant S-CSI, thus allocating more power to user $4$ is beneficial.

\section{Conclusions} \label{section_conclusion}
In this paper, we studied a new TTS-based joint active and passive beamforming optimization problem for an IRS-aided multiuser system. The weighted sum-rate was maximized under practical discrete phase-shift constraints at the IRS with only S-CSI. We proposed a novel TTS transmission protocol, where the long-term IRS phase shifts are optimized according to the S-CSI and the short-term transmit precoding vectors at the AP are designed adaptive to the instantaneous effective CSI with fixed phase shifts. A PDD-based algorithm and an SSCA algorithm were proposed for the single-user and multiuser cases, respectively. Simulation results showed that significant sum-rate performance gain can be achieved by using IRS based on S-CSI and with discrete phase shifters as compared to the case without IRS, especially when the deterministic Rician components dominate the channel and/or the channel correlation coefficients are large. It was also unveiled that channel correlations of the AP-IRS and IRS-user links exhibit distinct impacts on the proposed SSCA algorithm performance under different SNR regimes.

\begin{appendix}
	\subsection{Proof of Proposition \ref{prop_single_user}} \label{appendix_proof_proposition1}
	Given the spatial correlation matrix $\tilde{\bm{\Phi}}$ and the deterministic component $\mathbf{H}_{\textrm{LoS}}$, by plugging \eqref{optimal_w_single_user} into $\bar{r}$, the conditional achievable rate of the user is given by $\bar{r}(\bm{\Theta}|\tilde{\bm{\Phi}}, \mathbf{H}_{\textrm{LoS}}) 
	= \mathbb{E}[ \log_2 ( 1+P \|\mathbf{h}_r^H \bm{\Theta} \mathbf{G} + \mathbf{h}_d^H\|^2/\sigma^2)]$. 
	For a given power budget $P$, problem \eqref{TTS_problem_single_user} is equivalent to
	\begin{equation} \label{original_problem}
	\begin{aligned}
	\max\limits_{\bm{\Theta}} \; & \bar{r}(\bm{\Theta}|\tilde{\bm{\Phi}}, \mathbf{H}_{\textrm{LoS}}) \\
	\textrm{s.t.}\; & \phi_n \in \mathcal{F},\;\forall n \in \mathcal{N}.
	\end{aligned}
	\end{equation}
	
	Next, due to the concavity of the log function and according to Jensen's inequality, $\bar{r}(\bm{\Theta}|\tilde{\bm{\Phi}}, \mathbf{H}_{\textrm{LoS}})$ can be upper-bounded by $\bar{r}(\bm{\Theta}|\tilde{\bm{\Phi}}, \mathbf{H}_{\textrm{LoS}}) \leq \log_2(1+\mathbb{E}\{P \|\mathbf{h}_r^H \bm{\Theta} \mathbf{G} + \mathbf{h}_d^H\|^2/\sigma^2\})$.
	For $\mathbb{E}\{\|\mathbf{h}_r^H \bm{\Theta} \mathbf{G} + \mathbf{h}_d^H\|^2\}$, since $\mathbf{z}_r$, $\mathbf{F}$ and $\mathbf{z}_d$ are assumed to be independent of each other, we obtain
	\begin{equation} \label{stochastic_term_single_user}
	\begin{aligned}
	& \mathbb{E}\{  \|\mathbf{h}_r^H \bm{\Theta} \mathbf{G} + \mathbf{h}_d^H\|^2\} \\
	& =  \mathbb{E}\{\|(\bar{\mathbf{z}}_r^H +  \mathbf{z}_r^H\bm{\Phi}_{r,u}^{1/2H}) \bm{\Theta} (\bar{\mathbf{F}} + \bm{\Phi}_r^{1/2} \mathbf{F} \bm{\Phi}_d^{1/2}) \\
	 & \quad + \bar{\mathbf{z}}_d^H + \mathbf{z}_d^H \bm{\Phi}_d^{1/2H} \|^2 \}\\
	& = \|\mathbf{x}_1\|^2 + \mathbb{E}\{\| \mathbf{x}_2 \|^2\} + \mathbb{E}\{\| \mathbf{x}_3 \|^2\} \\
	& \quad + \mathbb{E}\{\| \mathbf{x}_4 \|^2\} + \mathbb{E}\{\| \mathbf{x}_5 \|^2\},
	\end{aligned}
	\end{equation}
	where $\bm{\Phi}_{r,u}$ denotes the spatial correlation matrix between IRS and the user, which is a Hermitian matrix \cite{McKay2005}, $\mathbf{x}_1 = \bar{\mathbf{z}}_r^H  \bm{\Theta} \bar{\mathbf{F}} + \bar{\mathbf{z}}_d^H$, $\mathbf{x}_2 = \bar{\mathbf{z}}_r^H \bm{\Theta}\bm{\Phi}_r^{1/2} \mathbf{F} \bm{\Phi}_d^{1/2}$, $\mathbf{x}_3 =  \mathbf{z}_r^H\bm{\Phi}_{r,u}^{1/2H}  \bm{\Theta}  \bar{\mathbf{F}}$, $\mathbf{x}_4 = \mathbf{z}_r^H\bm{\Phi}_{r,u}^{1/2H}  \bm{\Theta}  \bm{\Phi}_r^{1/2} \mathbf{F} \bm{\Phi}_d^{1/2}$, and $\mathbf{x}_5 = \mathbf{z}_d^H \bm{\Phi}_d^{1/2H}$.

	For $\| \mathbf{x}_1 \|^2$ in \eqref{stochastic_term_single_user}, we have
	\begin{equation} \label{x1}
	\begin{aligned}
	\| \mathbf{x}_1 \|^2 
	& =  \mathbf{v}^H \textrm{diag}\{\bar{\mathbf{z}}_r^H\} \bar{\mathbf{F}} \bar{\mathbf{F}}^H  \textrm{diag}\{\bar{\mathbf{z}}_r\}\mathbf{v} 
	+ \mathbf{v}^H \textrm{diag}\{\bar{\mathbf{z}}_r^H\}\bar{\mathbf{F}}  \bar{\mathbf{z}}_d \\
	& \quad + \bar{\mathbf{z}}_d^H \bar{\mathbf{F}}^H\textrm{diag}\{\bar{\mathbf{z}}_r\}\mathbf{v} + \bar{\mathbf{z}}_d^H \bar{\mathbf{z}}_d,
	\end{aligned}
	\end{equation}
	where we have applied the change of variables $\bar{\mathbf{z}}_r^H  \bm{\Theta} \bar{\mathbf{F}} = \mathbf{v}^H \textrm{diag}\{\bar{\mathbf{z}}_r^H\} \bar{\mathbf{F}}$. Then, we can derive the remaining terms in \eqref{stochastic_term_single_user} as follows. By expanding the terms in $\mathbb{E}\{\| \mathbf{x}_2 \|^2\}$, we have
	\begin{equation} \label{eq_1}
	\mathbb{E}\{\| \mathbf{x}_2 \|^2\} =  \mathbb{E}\{ \bar{\mathbf{z}}_r^H \bm{\Theta} \bm{\Phi}_r^{1/2} \mathbf{F} \bm{\Phi}_d^{1/2}  \bm{\Phi}_d^{1/2H} \mathbf{F}^H \bm{\Phi}_r^{1/2H} \bm{\Theta}^H \bar{\mathbf{z}}_r\}.
	\end{equation}
As $\bm{\Phi}_d^{1/2}$ and $\bm{\Phi}_r^{1/2}$ are Hermitian matrices, by letting $\bm{\Phi}_d \triangleq \mathbf{Q} \bm{\Sigma} \mathbf{Q}^H$ ($\mathbf{Q}$ is an unitary matrix), it can be shown that $\mathbb{E}\{\mathbf{F} \bm{\Phi}_d^{1/2}  \bm{\Phi}_d^{1/2H} \mathbf{F}^H\} = \mathbb{E}\{\mathbf{F} \mathbf{Q}\bm{\Sigma} \mathbf{Q}^H \mathbf{F}^H\} \overset{(a)}{=} \mathbb{E}\{\mathbf{F}\bm{\Sigma}\mathbf{F}^H\} \overset{(b)}{=} \big( \sum\nolimits_{i=1}^M\lambda_i \big)\frac{1}{1+\beta_{AI}}\mathbf{I}$, 
	where $(a)$ is due to the fact that multiplying an i.i.d. complex Gaussian matrix by a unitary matrix will not change its distribution, and $(b)$ is due to $\bm{\Sigma} = \textrm{diag} \{\lambda_1,\lambda_2,\cdots, \lambda_M\}$. Then, it follows that \eqref{eq_1} can be expressed as $ \mathbb{E}\{\| \mathbf{x}_2 \|^2\} 
	=  \frac{1}{1+\beta_{AI}}\big( \sum\nolimits_{i=1}^M\lambda_i \big) \mathbf{v}^H\textrm{diag}\{ \bar{\mathbf{z}}_r^H\} \bm{\Phi}_r \textrm{diag}\{\bar{\mathbf{z}}_r\} \mathbf{v}$. Similarly, we have
\begin{equation} \label{x345}
\begin{aligned}
	\mathbb{E}\{\|\mathbf{x}_3\|^2\} 
&	= \mathbb{E}\{\mathbf{v}^H \textrm{diag}\{\mathbf{z}_r^H \bm{\Phi}_{r,u}^{1/2H}\} \bar{\mathbf{F}} \bar{\mathbf{F}}^H  \textrm{diag}\{ \bm{\Phi}_{r,u}^{1/2} \mathbf{z}_r \}  \mathbf{v}\}\\
&  = \frac{1}{1+\beta_{Iu}} \mathbf{v}^H   (\bm{\Phi}_{r,u} \odot (\bar{\mathbf{F}} \bar{\mathbf{F}}^H)) \mathbf{v},\\
	\mathbb{E}\{\|\mathbf{x}_4\|^2\}
&	= \left( \sum\limits_{i=1}^M\lambda_i \right) \frac{1}{1+\beta_{AI}}\mathbb{E}\{ \mathbf{z}_r^H\bm{\Phi}_{r,u}^{1/2H}  \bm{\Theta}  \bm{\Phi}_r \bm{\Theta}^H \bm{\Phi}_{r,u}^{1/2} \mathbf{z}_r   \} \\
	& = \left( \sum\limits_{i=1}^M\lambda_i \right) \frac{1}{(1+\beta_{AI})(1+\beta_{Iu})} \mathbf{v}^H( \bm{\Phi}_{r,u}\odot \bm{\Phi}_r) \mathbf{v},\\
\mathbb{E}\{\|\mathbf{x}_5\|^2\} & = \mathbb{E}\{ \mathbf{z}_d^H \bm{\Phi}_d\mathbf{z}_d\} = \frac{1}{1+\beta_{Au}} \sum\limits_{i=1}^M \bm{\Phi}_d(i,i).
\end{aligned}
\end{equation}
	
Therefore, by combining the results in \eqref{x1}, \eqref{eq_1}, \eqref{x345} and extracting the terms that involve $\mathbf{v}$ (or equivalently $\bm{\Theta}$), we can approximate  problem \eqref{original_problem} by problem \eqref{statistic_problem}, which thus completes the proof.

	\subsection{Optimal Solution under $\beta_{AI} =\beta_{Iu} = 0$} \label{appendix_proof_proposition2}
	If $\beta_{AI} = 0$ and $\beta_{Iu} = 0$, problem \eqref{statistic_problem} reduces to
	\begin{equation} \label{statistic_problem_reduce}
	\begin{aligned}
	\max\limits_{\mathbf{v}} \; & \mathbf{v}^H ( \bm{\Phi}_{r,u} \odot \bm{\Phi}_r)\mathbf{v} \\
	\textrm{s.t.}\; & v_n \in \mathcal{F},\;\forall n \in \mathcal{N}.
	\end{aligned}
	\end{equation}
	This suggests that if both AP-IRS and IRS-user channels are Rayleigh fading, then the IRS phase shifts are only related to the correlation matrix $\bm{\Phi}_{r,u}$ and $\bm{\Phi}_r$. As a result, since the correlation coefficients in $\bm{\Phi}_{r,u}$ and $\bm{\Phi}_r$ are assumed to be non-negative real numbers, the elements in $\bm{\Phi}$ are all non-negative real numbers. Therefore, the objective function of problem \eqref{statistic_problem_reduce} satisfies $\sum\nolimits_{i\in\mathcal{N}} \sum\nolimits_{j\in\mathcal{N}} v_i^* \Phi(i,j)  v_j \leq \sum\nolimits_{i\in\mathcal{N}} \sum\nolimits_{j\in\mathcal{N}} \Phi(i,j) |v_i^*|  |v_j|$, 
	where $\Phi(i,j)$ denotes the element on the $i$-th row and $j$-th column of $\bm{\Phi}_{r,u} \odot \bm{\Phi}_r$ and the equality holds when $\angle v_i = \angle v_j$ and $|v_i| = |v_j|=1,\;\forall i,j$. In other words, the optimal solution of problem \eqref{statistic_problem_reduce} can be simply expressed as $\bm{\mathbf{v}} = \bar{\phi} \mathbf{1}$, for any arbitrary $\bar{\phi}$ that satisfies $|\bar{\phi}| = 1$. This thus completes the proof.

	\subsection{Jacobian Matrix of the Instantaneous Rate w.r.t. $\mathbf{v}^*$} \label{Jacobian_matrix}
For a given channel realization $\tilde{\mathbf{H}}$, the Jacobian matrix of the instantaneous rate vector $\mathbf{r}(\mathbf{v},\{\mathbf{w}_k\}; \tilde{\mathbf{H}})$ w.r.t. $\mathbf{v}^*$ is $\mathbf{J}_{\mathbf{v}^*}(\mathbf{v},\{\mathbf{w}_k\}; \tilde{\mathbf{H}}) = [ \nabla_{\mathbf{v}^*} r_1, \nabla_{\mathbf{v}^*} r_2, \cdots, \nabla_{\mathbf{v}^*} r_K]$. 
	According to the matrix calculus and the chain rule, it follows that
	\begin{equation}
	\nabla_{\mathbf{v}^*} r_k = \frac{\mathbf{a}_k}{\Gamma_k} - \frac{\mathbf{a}_{-k}}{\Gamma_{-k}},
	\end{equation} 
	where $\Gamma_k = \sum\nolimits_{j \in \mathcal{K}} |(\mathbf{v}^H \textrm{diag}(\mathbf{h}_{r,k}^H) \mathbf{G} + \mathbf{h}_{d,k}^H)\mathbf{w}_j|^2 + \sigma_k^2$, $\Gamma_{-k} = \sum\nolimits_{j \in \mathcal{K}\backslash k} |(\mathbf{v}^H \textrm{diag}(\mathbf{h}_{r,k}^H) \mathbf{G} + \mathbf{h}_{d,k}^H)\mathbf{w}_j|^2 + \sigma_k^2$, $ \mathbf{a}_k = \sum\nolimits_{j \in \mathcal{K}}\big(\textrm{diag}(\mathbf{h}_{r,k}^H)\mathbf{G} \mathbf{w}_j \mathbf{w}_j^H \mathbf{G}^H\textrm{diag}(\mathbf{h}_{r,k})\mathbf{v} +  \textrm{diag}(\mathbf{h}_{r,k}^H) \mathbf{G} \mathbf{w}_j  \mathbf{w}_j^H \mathbf{h}_{d,k} \big)$, and $\mathbf{a}_{-k} = \sum\nolimits_{j \in \mathcal{K}\backslash k} (\textrm{diag}(\mathbf{h}_{r,k}^H)\mathbf{G}$ $ \mathbf{w}_j \mathbf{w}_j^H \mathbf{G}^H\textrm{diag}(\mathbf{h}_{r,k})\mathbf{v} +  \textrm{diag}(\mathbf{h}_{r,k}^H)\mathbf{G} \mathbf{w}_j \mathbf{w}_j^H \mathbf{h}_{d,k} )$.
\end{appendix}

\bibliographystyle{IEEETran}
\scriptsize
\bibliography{references}

\end{document}